\documentclass[12pt]{article}

\usepackage{verbatim,color,amssymb}
\usepackage{amsmath}					
\usepackage{amsthm}					
\usepackage{natbib}
\usepackage{multirow}
\usepackage{setspace}
\usepackage[mathscr]{euscript}
\usepackage{fancyhdr}
\usepackage{enumitem}
\usepackage{graphicx}
\usepackage{geometry}
	\usepackage[bookmarks=false]{hyperref}
\usepackage{tabularx, booktabs}
\newcolumntype{Y}{>{\centering\arraybackslash}X}
\usepackage{lscape}

\usepackage{subfig}
\usepackage{float}
\usepackage{lineno}

\def\B{{\cal B}}
\def\C{{\cal C}}

\def\I{{\cal I}}

\def\M{{\cal M}}

\def\calP{{\cal P}}
\def\S{{\cal S}}

\def\diag{\hbox{diag}}

\def\wh{\widehat}
\def\wt{\widetilde}

\def\diag{\hbox{diag}}
\def\log{\hbox{log}}

\def\var{\hbox{var}}

\def\corr{\hbox{corr}}

\def\Bern{\hbox{Bernoulli}}
\def\Beta{\hbox{Beta}}

\def\Dir{\hbox{Dir}}

\def\Ga{\hbox{Ga}}

\def\MVN{\hbox{MVN}}

\def\Mult{\hbox{Mult}}

\def\P_25_ICML{{\it Proceedings of the 25th international conference on Machine learning}}

\def\refhg{\hangindent=20pt\hangafter=1}
\def\refmark{\par\vskip 2mm\noindent\refhg}

\def\refhg{\hangindent=20pt\hangafter=1}
\def\refmark{\par\vskip 2mm\noindent\refhg}

\def\bse{\begin{eqnarray*}}
\def\ese{\end{eqnarray*}}
\def\be{\begin{eqnarray}}
\def\ee{\end{eqnarray}}
\def\bq{\begin{equation}}
\def\eq{\end{equation}}

\def\wh{\widehat}

\def\trans{^{\rm T}}

\def\th{^{th}}

\def\Data{{\hbox{Data}}}

\def\b1e{{\mathbf e}}
\def\b1f{{\mathbf f}}

\def\bP{{\mathbf P}}

\def\bu{{\mathbf u}}

\def\bzero{{\mathbf 0}}

\newcommand{\bmu}{\mbox{\boldmath $\mu$}}
\newcommand{\bDelta}{\mbox{\boldmath $\Delta$}}

\newcommand{\bpi}{\mbox{\boldmath $\pi$}}

\newcommand{\bzeta}{\mbox{\boldmath $\zeta$}}

\newcommand{\bSigma}{\mbox{\boldmath $\Sigma$}}

\newcommand{\blambda}{\mbox{\boldmath $\lambda$}}

\newcommand{\abs}[1]{\left\vert#1\right\vert}

\renewcommand\footnoterule{\kern-3pt \hrule \textwidth 2in \kern 2.6pt}

\def\boxit#1{\vbox{\hrule\hbox{\vrule\kern6pt \vbox{\kern6pt \textcolor{blue}{#1}\kern6pt}\kern6pt\vrule}\hrule}}

\def\authorfootnote#1{{\let\thefootnote\relax\footnotetext{#1}}}

\allowdisplaybreaks

\begin{document}
\thispagestyle{empty}
\baselineskip=28pt

\begin{center}
{\LARGE{\bf Bayesian Semiparametric Mixed Effects Markov Chains 
}}
\end{center}
\baselineskip=12pt

\begin{center}
Abhra Sarkar\\
Department of Statistical Science, Duke University, Box 90251, Durham NC 27708-0251, USA\\
abhra.sarkar@duke.edu\\ 
\vskip 10pt 
Jonathan Chabout\\
Department of Neurobiology, Duke University, Durham, NC 27710, USA\\
jchabout.pro@gmail.com\\
\vskip 10pt 
Joshua Jones Macopson\\
Department of Neurobiology, Duke University, Durham, NC 27710, USA\\
joshua.jones.macopson@duke.edu\\
\vskip 10pt 
Erich D. Jarvis\\
Department of Neurobiology, Duke University, Durham, NC 27710, USA\\
Howard Hughes Medical Institute, Chevy Chase, MD 20815, USA\\
The Rockefeller University, New York, NY 10065, USA\\
jarvis@neuro.duke.edu\\
\vskip 10pt 
David B. Dunson\\
Department of Statistical Science, Duke University, Box 90251, Durham NC 27708-0251, USA\\
dunson@duke.edu\\
\end{center}

\vskip 4mm
\begin{center}
{\Large{\bf Abstract}} 
\end{center}
\baselineskip=12pt

Studying the neurological, genetic and evolutionary basis of human vocal communication mechanisms 
using animal vocalization models is an important field of neuroscience. 
The data sets typically comprise structured sequences of syllables or `songs' produced by animals from different genotypes under different social contexts. 
We develop a novel Bayesian semiparametric framework for inference in such data sets. 
Our approach is built on a novel class of mixed effects Markov transition models for the songs 
that accommodates exogenous influences of genotype and context as well as animal specific heterogeneity.
We design efficient Markov chain Monte Carlo algorithms for posterior computation. 
Crucial advantages of the proposed approach include its ability to provide insights into key scientific queries related 
to global and local influences of the exogenous predictors on the transition dynamics via automated tests of hypotheses. 
The methodology is illustrated using simulation experiments and the aforementioned motivating application in neuroscience.

\vskip 8mm
\baselineskip=12pt
\noindent\underline{\bf Some Key Words}: Bayesian nonparametrics, Categorical sequences, Markov chains, Mixed effects models, Mouse vocalization experiments. 

\par\medskip\noindent
\underline{\bf Short Title}: Mixed Effects Markov Chains 

\par\medskip\noindent

\pagenumbering{arabic}
\setcounter{page}{0}
\newlength{\gnat}
\setlength{\gnat}{16pt}
\baselineskip=\gnat

\newpage

\section{Introduction}

This article introduces a novel class of mixed effects Markov models 
for categorical sequences recorded under the influence of exogenous categorical predictors.  
The predictors themselves do not vary sequentially but remain fixed for the entire lengths of the sequences. 
Additionally, each sequence may be associated with an individual selected at random from a larger population of interest 
and may thus be influenced by random effects specific to the chosen subjects. 
Our inferential goals include estimation of transition probabilities governing the stochastic evolution of the sequences
as well as an assessment of the importance of the predictors in influencing the transition dynamics. 

While the literature on Markov models and its various derivatives and extensions is enormous, 
owing to numerous methodological and computational challenges, 
the literature on mixed effects Markov models is relatively sparse in spite of their immense potential and general practical utility. 

Generalized linear models (GLM) provide one route to modeling mixed effects Markov chains.  
However, except for binary sequences, specifying flexible predictor-dependent Markov models using traditional GLMs is daunting.   
Consider, for example, an adaptation of the multinomial logit model \citep{mccullagh_nelder:1989,agresti:2013} to our setting. 
With $p$ categorical predictors $x_{j}$ with $d_{j}$ categories each, $j=1,\dots,p$,  
and response sequences $y_{t}$ comprising $d_{0}$ categories, 
even without any interaction or random effects, 
such a specification would require formulating $d_{0}-1$ models, one for each $y_{t}=1,\dots,d_{0}-1$, 
with $\sum_{j=0}^{p}(d_{j}-1)$ dummy variables representing local dependence and predictors $y_{t-1},x_{1},\dots,x_{p}$
included as linear predictors on the logit scale. 
Interactions and random effects greatly add to the complexity; for example, due 
to the lack of analytic forms marginalizing out the random effects. 
Inferences under such models and testing of hypotheses become intractable. 
Hence, relevant work in the literature has 
focused on very simple scenarios, such as binary sequences, single predictors, and models excluding interactions and
random effects \citep{Bonney:1987,fitzmaurice_liard:1993,Azzalini:1994,schildcrout_heagerty:2005,schildcrout_heagerty:2007,rahman_islam:2007,bizzotto_etal:2010,islam_etal:2013}.

Markov chains provide building blocks for many other important dynamical systems, including hidden Markov models (HMM). 
HMMs consist of a latent process $y_{t}$, evolving according to a Markov chain, and an observed process $z_{t}$, evolving independently, given $y_{t}$. 
\cite{hudges_guttorp:1994} and \cite{spezia:2006} developed fixed effects HMMs with sequentially varying predictors influencing the transition dynamics of the latent sequence. 
\cite{turner_etal:1998} and \cite{wang_puterman:1999} discussed mixed effects HMMs for count data, 
but incorporated the covariate effects only through the observed process. 
See also \cite{maruotti_ryden:2009,delattre:2010,maruotti:2011} and \cite{rueda2013bayesian}. 
\cite{altman:2007} developed a mixed HMM, adopting a multinomial logit model based approach to incorporate mixed effects in the transition distributions, 
but focused only on models with binary sequences with a single binary predictor in real and simulated illustrations. 
Additional discussions on these models are presented in the Supplementary Materials.

This article presents a fundamentally different approach to modeling mixed effects Markov chains directly 
motivated by mouse vocalization experiments. 
Our proposed Bayesian hierarchical formulation provides flexible, easily interpretable representations of 
exogenous predictor dependent individual specific transition probability matrices, borrowing information across sequences via a layered hierarchy.  
We include fixed and random effects directly on the probability scale, avoiding issues in choosing link functions and facilitating computational
simplifications. 
Using Dirichlet distributed random effects, we obtain analytic forms for the population level probabilities. 
To reduce dimensionality, the model clusters levels of predictors that have similar impact on the transition dynamics.  
The model structure leads to a stable and 
efficient Markov chain Monte Carlo (MCMC) algorithm. 
Estimates of the posterior probabilities of global and local hypotheses are 
easily available from samples drawn by the algorithm.

The article is organized as follows. 
Section \ref{sec: mve} describes the neuroscience applications that motivated our research. 
Section \ref{sec: memc} develops the model. 
Section \ref{sec: pi} describes MCMC algorithms to sample from the posterior.
Section \ref{sec: afds} presents the results of the proposed method applied to mouse vocalization data. 
Section \ref{sec: se} presents the results of simulation experiments.
Section \ref{sec: dscsn} contains concluding remarks.

\section{Mouse Vocalization Experiments} \label{sec: mve}

While the statistical problem that we address is of broad scope, 
our research was motivated by studies of the genetic and evolutionary basis of human vocal communication. 
Spoken language plays a central role in human culture. 
We belong to one of few species that can learn to produce new vocalizations, which we use to express emotions,
convey ideas, and communicate. 
These vocal behaviors are susceptible to a range of impairments, 
making dramatic impacts on everyday life 
and presenting a major public health issue, 
with the prevalence of speech-sound disorders in young children estimated at 8-9 percent 
\citep{NIH-NIDCD:2010}.
Such disorders are highly heritable, but causes are typically unknown \citep{Law_etal:2000}.

Many species of birds have the ability to sing complex
songs to communicate information within and across species \citep{marler_slabbekoorn:2004}. 
Songbirds are vocal learners like us and their songs are similar in many ways to human languages.    
Extensive research spanning over the last six decades has shown
that songbirds provide an immensely useful model 
for studying the neuroscience of human vocal communication systems \citep{zeigler_marler:2008}. 
Aside the human aspect, studying birds provides insights into the evolution of vocal communication skills across species \citep{kaas:2006}. 

\begin{figure}[h!]
\centering
\includegraphics[height=4cm, width=7cm, trim=0cm 0.25cm 0cm 0cm, clip=true]{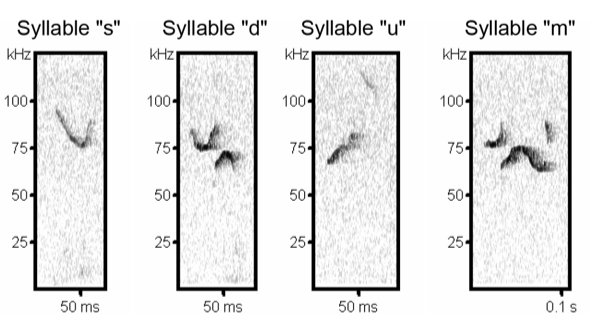}
\caption{Syllables making up mouse vocalizations.
}
\label{fig: syllables}
\end{figure}

\cite{Holy_Guo:2005} discovered that male mice `sing' ultrasonic vocalizations (USVs) with some features similar to courtship songs of songbirds. 
Being mammals, mice have a lot more in common with humans in terms of genetic makeup, anatomy and physiology.
Mice are also easier to manipulate genetically and less expensive to raise in controlled laboratory environments. 
In recent years, neuroscientists have thus invested tremendous efforts in studying the mouse vocalization system. 
Studies have confirmed that male mice emit USV songs in sexual and other contexts using a multisyllabic repertoire. 
The syllables can be categorized based on spectral features \citep{Holy_Guo:2005,Scattoni_etal:2008,Arriaga_etal:2012}. 
\cite{Chabout_etal:2012,Chabout_etal:2015} grouped syllables into four categories - 
simple syllables without any pitch jumps (`s'); 
complex syllables containing two notes separated by a single upward (`u') or downward (`d') pitch jump; 
and more complex syllables containing a series of multiple pitch jumps (`m'). 
See Figure \ref{fig: syllables}. 
Syllable durations and inter-syllable intervals are typically between 0 and 250 ms 
\citep{Chabout_etal:2015,Castellucci_etal:2016}.  
Each interval of 250ms that a mouse remains silent is thus classified as a special silence syllable (`x'). 


\begin{figure}[h!]
\centering
\includegraphics[height=3cm, width=15.5cm, trim=0cm 11.75cm 0cm 11.75cm, clip=true]{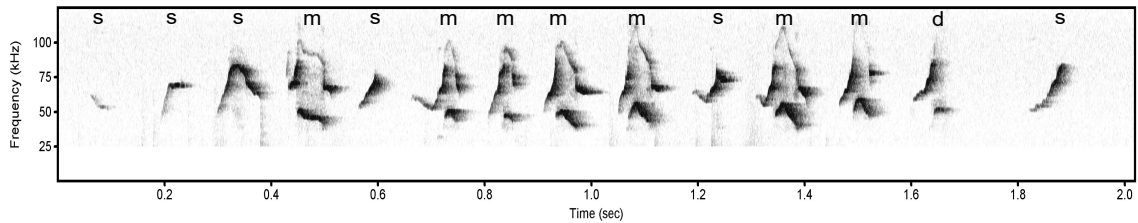}
\vspace{-0.5cm}
\caption{An example song segment sung by a wild type mouse under the U context.
}
\label{fig: example song}
\end{figure}

The influences of social stimuli and genetic differences on mouse song syntax are also being extensively investigated. 
\cite{Chabout_etal:2012} studied the role of USVs in triggering and maintaining social interaction between male mice. 
%
%
\cite{Musolf_etal:2015} showed that the repertoire of male mice USVs differ among strains. 
\cite{Scattoni_etal:2008} compared the USV repertoire in a strain of mouse 
that displays social abnormalities and repetitive behaviors analogous to symptoms of autism, 
with USVs produced by three other control strains. 
\cite{Chabout_etal:2015} 
studied if male mice change their syntax according to social contexts, exposing them to four different stimuli - 
fresh female urine 
on a cotton tip placed inside the male's cage (U), 
awake and behaving adult female placed inside the cage (L), 
an anesthetized female placed on the lid of the cage (A), 
and anesthetized male placed on the lid of the cage (AM). 

\cite{Chabout_etal:2016} investigated the effect of a rare mutation of the Foxp2 (forkhead-box P2) gene on the mouse vocalization patterns. 
Many human spoken language disorders are suspected to be caused by this mutation. 
Affected individuals have difficulties mastering the coordinated movement sequences of
syllables that underlie fluent speech, described as developmental verbal dyspraxia
(DVD) or childhood apraxia of speech (CAS), accompanied by other problems with verbal and written language. 
%
The Foxp2 orthologs' coding sequences and brain expressions are remarkably conserved across species, including humans and mice \citep{Lai_etal:2003,Haesler_etal:2004}. 
It is thus interesting to explore if the mouse vocalization model can be used to gain insights 
into the role of Foxp2 in vocalization capabilities across the two mammalian species. 
To this end, 
\cite{Chabout_etal:2016} compared songs sung by mice carrying the mutation with songs produced by wild type mice 
under the female mouse related stimuli U, L and A. 
Other mouse vocalization experiments studying the role of Foxp2 include \cite{Fujita_etal:2008,Castellucci_etal:2016,Gaub_etal:2016} etc.

Notably, each of these data sets can be described as a collection of songs recorded under different combination of values of associated predictors. 
The data sets studied in \cite{Scattoni_etal:2008,Fujita_etal:2008}, for example, comprised songs produced by mice from different genotypes. 
The data sets studied in \cite{Chabout_etal:2012,Chabout_etal:2015} consisted of songs recorded under different social stimuli. 
The data sets studied in \cite{Gaub_etal:2016,Chabout_etal:2016} comprised songs recorded under different combinations of genotype and context. 
Also, in all these studies, neuroscientists are primarily interested in assessing the global and local influences of 
the predictors on the mouse vocalization patterns and abilities.

\begin{table}[ht]
\begin{tabularx}{0.45\textwidth}{|c *{4}{|Y}|}
\multicolumn{4}{c}{Foxp2 Mutant} \\ \cline{1-4}
\multirow{2}{*}{Mouse ID} & \multicolumn{3}{c|}{Social Context} \\ \cline{2-4}
& U & L & A\\ \hline \hline 
F1 & 1423 & 4514 & 2107 \\ 
F2 & 3741 & 4099 & 4639 \\ 
F3 & 2516 & 5059 & 4610 \\ 
F4 & 3812 & 3207 & 2641 \\ 
F5 & 3947 & 4598 & 1833 \\ 
F6 & 730 & 3812 & 4479 \\ 
F7 & 3909 & 4806 & 2391 \\ 
F8 & 3347 & 3482 & 3335 \\ 
\hline
\end{tabularx}
\quad~~~
\begin{tabularx}{0.45\textwidth}{|c *{4}{|Y}|}
\multicolumn{4}{c}{Wild Type} \\ \cline{1-4}
\multirow{2}{*}{Mouse ID} & \multicolumn{3}{c|}{Social Context} \\ \cline{2-4}
& U & L & A\\ \hline \hline 
W1 & 3787 & 6063 & 5246 \\ 
W2 & 4189 & 4975 & 1540 \\ 
W3 & 3453 & 4971 & 5085 \\ 
W4 & 2460 & 1776 & 3771 \\ 
W5 & 2817 & 4806 & 3445 \\ 
W6 & 2489 & 4281 & 587 \\ 
\hline
\end{tabularx}
\caption{General structure of the Foxp2 data set studied in \cite{Chabout_etal:2016}.
Each cell represents the number of syllables making up the corresponding song. }
\label{tab: Foxp2 data set}
\end{table}

While the literature on mouse vocalization studies has already become substantial and is growing very rapidly, 
statistical methods for analysis of syntax differences have lagged behind. 
Neuroscientists have thus often focused only on repertoire differences, 
comparing the overall proportions of syllables between different combinations of exogenous predictors, 
but have largely ignored the problem of studying systematic differences in sequential arrangements of these syllables within songs 
\citep{Scattoni_etal:2008,Chabout_etal:2012,Musolf_etal:2015,Gaub_etal:2016}.

The mixed effects Markov model proposed in this article aims to provide a sophisticated statistical framework for assessing
syntax differences due to exogenous factors. 
While the methodology is readily applicable to all the studies described above,  
to illustrate its utility, we will focus specifically on the Foxp2 study reported in \cite{Chabout_etal:2016}. 
%
%
The general structure of the Foxp2 data set is depicted in Table \ref{tab: Foxp2 data set} 
where $42$ songs of various lengths sung by $8$ Foxp2 mutant mice 
and $6$ wild type mice were recorded under the contexts U, L and A. 
Independent analyses of the songs using higher order models \citep{sarkar_dunson:2016} were strongly indicative of first order Markovian dynamics. 
The hypothesis of primary scientific interest postulates the Foxp2 mutation to significantly affect the mouse vocalization patterns and abilities across all contexts. 
A hypothesis of secondary interest is that the vocalization patterns also vary significantly between contexts across genotypes. 
Additionally, if the global effect of the Foxp2 mutation turns out to be significant, 
we are interested in assessing how the mutation affects the vocalization patterns locally for each fixed social context. 

\section{Mixed Effects Markov Chains}\label{sec: memc}

Consider a collection of categorical sequences $\{y_{s,t}\}_{s=1,t=1}^{s_{0},T_{s}}$ with $y_{s,t} \in \S_{0}=\{1\dots,d_{0}\}$.
Each sequence $s$ has, associated with it, $p$ exogenous categorical predictors $\{x_{s,j}\}_{j=1}^{p}$ with $x_{s,j}\in \S_{j}=\{1,\dots,d_{j}\}$.   
These exogenous predictors remain fixed over time for each sequence. 
The notations $y_{t}, x_{j}$ sans the subscript $s$ will sometimes be used  to refer collectively to $y_{s,t}$, $x_{s,j}$ and also to denote values taken by them. 
The probability of $y_{s,t}$ taking a value in $\S_{0}$ is assumed to depend on its immediate previous value $y_{s,t-1}$  
and possibly also on the values taken by the associated exogenous predictors $\{x_{s,j}\}_{j=1}^{p}$, 
and is denoted by $P_{x_{s,1},\dots,x_{s,p}}(y_{s,t} \mid y_{s,t-1})$.  
We refer to such sequences as Markov Chains with Exogenous Predictors (MCEP). 
The dynamics of the $s\th$ sequence is thus governed by the model  
\bse
&& (y_{s,t} \mid y_{s,t-1}, x_{s,j}, j=1,\dots,p)  \sim \nonumber\\
&& \hspace{1cm}  \Mult[\{1,\dots,d_{0}\},P_{x_{s,1},\dots,x_{s,p}}(1\mid y_{s,t-1}),\dots,P_{x_{s,1},\dots,x_{s,p}}(d_{0}\mid y_{s,t-1})].  \label{eq: mcep}
\ese

Additionally, 
each sequence $s$ may also be associated with an individual $i_{s}=i$
drawn at random from a larger population of interest $\I$. 
The individual level transition probabilities, denoted by $P_{x_{1},\dots,x_{p}}^{(i)}(y_{t} \mid y_{t-1})$, 
may then be assumed to have been randomly drawn from a population of transition probabilities 
$\calP=\{P_{x_{1},\dots,x_{p}}^{(i)}(y_{t} \mid y_{t-1}): i \in \I\}$ 
with mean transition probability parameters $P_{x_{1},\dots,x_{p}}(y_{t} \mid y_{t-1})$. 
The model for the process governing the evolution of the $s\th$ sequence is now modified as 
\bse
&& (y_{s,t} \mid y_{s,t-1}, x_{s,j}, j=1,\dots,p,i_{s})  \sim \nonumber\\
&& \hspace{1cm}  \Mult[\{1,\dots,d_{0}\},P_{x_{s,1},\dots,x_{s,p}}^{(i_{s})}(1\mid y_{s,t-1}),\dots,P_{x_{s,1},\dots,x_{s,p}}^{(i_{s})}(d_{0}\mid y_{s,t-1})].  \label{eq: memm}
\ese

Key inferential goals include estimation of individual and population level transition probabilities, 
$P_{x_{1},\dots,x_{p}}^{(i)}(y_{t} \mid y_{t-1})$ and $P_{x_{1},\dots,x_{p}}(y_{t} \mid y_{t-1})$, 
and an assessment of the global and local influence of the predictors on these transition probabilities. 
The hypotheses of a global impact of the $j\th$ predictor correspond to 
\bse
&& H_{0j}: P_{x_{1},\dots,x_{p}}^{(i)}(y_{t} \mid y_{t-1})~\text{does not vary with values of}~x_{j}~~~\text{vs} \nonumber\\
&& H_{1j}: P_{x_{1},\dots,x_{p}}^{(i)}(y_{t} \mid y_{t-1})~\text{varies with varying values of}~x_{j}.
\ese
Local hypotheses instead correspond to pairwise comparisons between transition probabilities for two 
different levels of $x_j$.

In our motivating application, $i_s$ indexes the mouse under study, $y_{s,t}\in\S_{0}=\{d,m,s,u,x\}=\{1,2,3,4,5\}$ denotes
the sequence of `syllables' measuring the song dynamics, and there are two categorical predictors - 
genotype $x_{s,1}\in\S_{1}=\{F,W\}=\{1,2\}$ and context $x_{s,2}\in\S_{2}=\{U,L,A\}=\{1,2,3\}$. 
A key interest is in studying the impact of a mutation in the Foxp2 gene on vocalization patterns across different contexts.  

Towards these goals, we first focus, 
in Section \ref{sec: pm for mcep}, 
on the simpler problem of developing a statistical framework for MCEPs 
that facilitates testing the significance of the exogenous predictors, ignoring any individual specific effect.
The problem of modeling individual level transition probabilities incorporating individual specific random effects is then addressed in Section \ref{sec: memcep}.

\subsection{A Partition Model for MCEP} \label{sec: pm for mcep}

To model MCEPs, 
we construct a probabilistic partition of each $\S_{j}$ 
such that the values of $x_{j}$ that are clustered together have similar influences on the dynamics of $y_{s,t}$.   
Specifically, given a random partition $\C^{(j)}=\{\C_{\ell}^{(j)}\}$ of $\S_{j}, j=1,\dots,p$, 
we model the stochastic evolution of $y_{s,t}$ as
\begin{eqnarray}
&& (y_{s,t} \mid y_{s,t-1}, x_{s,j} \in \C_{\ell_{j}}^{(j)}, j=1,\dots,p)  \sim \nonumber\\
&& \hspace{1cm}  \Mult[\{1,\dots,d_{0}\},\lambda_{\ell_{1},\dots,\ell_{p}}^{\star}(1\mid y_{s,t-1}),\dots,\lambda_{\ell_{1},\dots,\ell_{p}}^{\star}(d_{0}\mid y_{s,t-1})],  \label{eq: pm1}
\end{eqnarray}
where $\blambda_{\ell_{1},\dots,\ell_{p}}^{\star}(\cdot\mid y_{t-1})=\{\lambda_{\ell_{1},\dots,\ell_{p}}^{\star}(1\mid y_{t-1}),\dots,\lambda_{\ell_{1},\dots,\ell_{p}}^{\star}(d_{0} \mid y_{t-1})\}\trans$ are probability vectors 
for each combination $(y_{t-1},\ell_{1},\dots,\ell_{p})$. 
Clearly, $1 \leq \abs{\C^{(j)}} \leq d_{j}$. 
When $\abs{\C^{(j)}}=1$, all $d_{j}$ categories are clustered together, 
and hence the evolution of $y_{s,t}$ does not depend on the specific value taken by the associated $j\th$ predictor $x_{s,j}$.  We will refer to this case as the null model $\M_{0j}$ 
corresponding to null hypothesis $H_{0j}$. At the other extreme, $\abs{\C^{(j)}}=d_{j}$ so that each of the $d_{j}$ categories of $x_{j}$ has its own cluster. 

Introducing latent cluster allocation variables, we can re-express (\ref{eq: pm1}) as 
\begin{eqnarray}
&& (y_{s,t} \mid y_{s,t-1},z_{j,x_{s,j}}=h_{j}, j=1,\dots,p)  \sim \nonumber\\
&& \hspace{1cm}  \Mult[\{1,\dots,d_{0}\}, \lambda_{h_{1},\dots,h_{p}}(1\mid y_{s,t-1}),\dots,\lambda_{h_{1},\dots,h_{p}}(d_{0}\mid y_{s,t-1})], \nonumber \\
&& z_{j,\ell} \sim \Mult[\{1,\dots,k_{j}\},\pi_{1}^{(j)},\dots,\pi_{k_{j}}^{(j)}], \label{eq: pm2}
\end{eqnarray}
where $\blambda_{h_{1},\dots,h_{p}}(\cdot\mid y_{t-1})=\{\lambda_{h_{1},\dots,h_{p}}(1\mid y_{t-1}),\dots,\lambda_{h_{1},\dots,h_{p}}(d_{0} \mid y_{t-1})\}\trans$ are probability vectors 
for each combination $(y_{t-1},h_{1},\dots,h_{p})$, 
and $\{z_{j,\ell}\}_{j=1,\ell=1}^{p,d_{j}}$ index allocation of the $d_{j}$ observed categories of the $j\th$ predictor to $k_{j}$ latent categories, 
inducing a partition $\C^{(j)}$ of $\S_{j}$. 
Two categories $\ell_{1},\ell_{2} \in \{1,\dots,d_{j}\}$ will be clustered together iff $z_{j,\ell_{1}}=z_{j,\ell_{2}}$. 

The above formulation allows the partition $\C^{(j)}$ to comprise at most $k_{j}$ sets. 
By allowing empty latent components, such hierarchical formulations define many-to-one mappings 
from the space of possible combinations of $z_{j,\ell}$ to the space of possible partitions. 
For example, with $d_{j}=4$ and $k_{j}=3$, $(z_{j,1},\dots,z_{j,4})=(1,1,1,3)$, $(z_{j,1},\dots,z_{j,4})=(2,2,2,1)$, $(z_{j,1},\dots,z_{j,4})=(3,3,3,2)$ 
all induce the same partition $\C^{(j)}=\{\{1,2,3\},\{4\}\}$ of $\{1,\dots,d_{j}\}$. 
Given $\{z_{j,\ell}\}_{\ell=1,j=1}^{d_{j},p}$ and the labeling schemes for the sets in the induced partitions $\{\C^{(j)}\}_{j=1}^{p}$, 
the $\blambda_{\ell_{1},\dots,\ell_{p}}^{\star}$'s and the $\blambda_{h_{1},\dots,h_{p}}$'s that are associated with the nonempty components are uniquely related.
For example, with $p=2$, $(z_{1,1},\dots,z_{1,4})=(1,1,1,3)$ and $(z_{2,1},\dots,z_{2,3})=(2,2,1)$, 
inducing the partitions $\C^{(1)}=\{\C_{1}^{(1)},\C_{2}^{(1)}\}=\{\{1,2,3\},\{4\}\}$ and $\C^{(2)}=\{\C_{1}^{(2)},\C_{2}^{(2)}\}=\{\{1,2\},\{3\}\}$, 
$\blambda_{1,1}^{\star}=\blambda_{1,2}$, $\blambda_{1,2}^{\star}=\blambda_{1,1}$, $\blambda_{2,1}^{\star}=\blambda_{3,2}$ and $\blambda_{2,2}^{\star}=\blambda_{3,1}$.

Model specification is completed with priors for probability vectors $\blambda_{h_{1},\dots,h_{p}}(\cdot\mid y_{s,t-1})$ 
and $\bpi^{(j)}=\{\pi_{1}^{(j)},\dots,\pi_{k_{j}}^{(j)}\}\trans$. 
It is natural to center the prior for the transition probability vectors $\blambda_{h_{1},\dots,h_{p}}(\cdot\mid y_{t-1})$ around some $\blambda_{0}(\cdot\mid y_{t-1})$,
which captures the overall transition dynamics. 
In addition, some states in $\C^{(0)}$ may be preferred to others across levels of the predictors.
Let $\blambda_{00}=\{\lambda_{00}(1),\dots,\lambda_{00}(d_{0})\}\trans$ be a probability vector capturing such global behaviors. 
A natural prior for $\blambda_{h_{1},\dots,h_{p}}(\cdot\mid y_{t-1})$ 
that allows sharing of information across different layers of hierarchy may be specified as  
\bse
&& \hspace{-2cm} \blambda_{h_{1},\dots,h_{p}}(\cdot\mid y_{t-1}) \sim \Dir\{\alpha_{0}\lambda_{0}(1\mid y_{t-1}),\dots,\alpha_{0}\lambda_{0}(d_{0}\mid y_{t-1})\},\\
&& \hspace{-2cm}  \blambda_{0}(\cdot\mid y_{t-1}) \sim \Dir\{\alpha_{00}\lambda_{00}(1),\dots,\alpha_{00}\lambda_{00}(d_{0})\},~~~~~~\alpha_{0}\sim\Ga(a_{\alpha_{0}},b_{\alpha_{0}}).
\ese
Likewise, a natural choice for the prior on $\bpi^{(j)}$ is given by
$\bpi^{(j)} \sim \Dir(\alpha_{j},\dots,\alpha_{j}).$ 
Marginalizing out $\bpi^{(j)}$, the induced prior on $\C^{(j)}=\{\C_{1}^{(j)},\dots,\C_{\wt{k}_{j}}^{(j)}\}$ is given by 
\vspace{-3ex}
\bse
p_{0}(\C^{(j)}) = \frac{{k_{j}}_{\left(\abs{\C^{(j)}}\right)}}{(k_{j}\alpha_{j})^{(d_{j})}} \prod_{\ell_{j}=1}^{\abs{\C^{(j)}}}\alpha_{j}^{\left(\abs{\C_{\ell_{j}}^{(j)}}\right)},
\ese
\vspace{-3ex}\\
where $x^{(i)}=x(x+1)\cdots(x+m-1)$ and $x_{(m)}=x(x-1)\cdots(x-m+1)$. 
%
The prior probability of the null model $\M_{0j}$ is then obtained as 
\bse
p_{0}(\M_{0j}) = p_{0}(\wt{k}_{j}=1) = {k_{j}  \alpha_{j}^{(d_{j})}} / {(k_{j}\alpha_{j})^{(d_{j})}}.
\ese
Since $1 \leq \abs{\C^{(j)}} \leq d_{j}$, a nonparametric partition model is obtained by setting $k_{j} = d_{j}$, 
with $\alpha_{j}=\alpha/k_{j}$, which converges weakly to a Dirichlet process as $k_j \to \infty$. 

Our proposed model is related to the rich literature on reducing dimensionality in characterizing predictor effects via partitioning; refer, for example to \cite{breiman1984classification,denison_etal:1998}, though we are not aware of such methods being developed in the setting of characterizing predictor effects 
on Markov dynamics.

\subsection{Mixed Effects MCEP}\label{sec: memcep}
Having developed a general framework for MCEP, 
we now address the problem of incorporating individual effects into the model. 
For ease of exposition and interpretation, we focus initially on settings similar to our motivating application of mouse vocalization experiments. 
Extensions to similar other scenarios are straightforward. 

In our motivating application, we have $m=1,\dots,m_{x_{1}}$ mice from genotypes $x_{1}\in\S_{1}=\{1,\dots,d_{1}\}$ 
singing under contexts $x_{2}\in\S_{2}=\{1,\dots,d_{2}\}$,
generating a total of $s_{0}=d_{2}\sum_{x_{1}=1}^{d_{1}}m_{x_{1}}$ songs $\{y_{s,t}\}_{s=1,t=1}^{s_{0},T_{s}}$.  
The model developed in Section \ref{sec: pm for mcep} assumes that under a given context all $m_{x_{1}}$ mice from the $x_{1}\th$ genotype sing according to the same probability model.  Introducing mouse-specific transition distributions $\blambda^{(i)}(\cdot\mid y_{t-1})=\{\lambda^{(i)}(1\mid y_{t-1}),\dots,\lambda^{(i)}(d_{0}\mid y_{t-1})\}\trans$, we characterize the transition dynamics of the different mice according to the hierarchical model
\begin{eqnarray}
&&(y_{s,t} \mid y_{s,t-1}, i_{s}=i, z_{j,x_{s,j}} =h_{j}, j=1,2)  \sim \nonumber\\
&& \hspace{1cm} \Mult[\{1,\dots,d_{0}\},P_{h_{1},h_{2}}^{(i)}(1\mid y_{s,t-1}),\dots,P_{h_{1},h_{2}}^{(i)}(d_{0}\mid y_{s,t-1})], ~~\text{where} \nonumber\\
&& \bP_{h_{1},h_{2}}^{(i)} (\cdot\mid y_{t-1}) = \pi_{0}(y_{t-1}) \blambda_{h_{1},h_{2}}(\cdot\mid y_{t-1}) + \pi_{1}(y_{t-1}) \blambda^{(i)}(\cdot\mid y_{t-1}),\nonumber\\
&& z_{j,\ell} \sim \Mult[\{1,\dots,d_{j}\},\pi_{1}^{(j)},\dots,\pi_{d_{j}}^{(j)}], ~~~~~~~~~~~ \bpi^{(j)} \sim \Dir(\alpha_{j},\dots,\alpha_{j}), \nonumber\\
&& \blambda^{(i)}(\cdot \mid y_{t-1}) \sim \Dir\{\alpha^{(0)}\lambda_{0}(1 \mid y_{t-1}),\dots,\alpha^{(0)}\lambda_{0}(d_{0} \mid y_{t-1})\}, \nonumber\\
&& \blambda_{h_{1},h_{2}}(\cdot\mid y_{t-1}) \sim \Dir\{\alpha_{0}\lambda_{0}(1 \mid y_{t-1}),\dots,\alpha_{0}\lambda_{0}(d_{0} \mid y_{t-1})\},\nonumber \\
&& \blambda_{0}(\cdot\mid y_{t-1}) \sim \Dir\{\alpha_{00}\lambda_{00}(1),\dots,\alpha_{00}\lambda_{00}(d_{0})\}, \nonumber \\
&& \pi_{0}(y_{t-1}) \sim \Beta(a_{0},a_{1}),~~~~~~\alpha_{0}\sim\Ga(a_{\alpha_{0}},b_{\alpha_{0}}),~~~~~~~\alpha^{(0)}\sim\Ga(a_{\alpha^{(0)}},b_{\alpha^{(0)}}).
\label{eq:MEMCEP}
\end{eqnarray}

Expression (\ref{eq:MEMCEP}) characterizes the mouse-specific transition probability $\bP_{h_{1},h_{2}}^{(i)} (\cdot\mid y_{t-1})$ as 
a convex combination of a {\em baseline} probability $\blambda_{h_{1},h_{2}}(\cdot\mid y_{t-1})$ and a mouse-specific 
random effect $\blambda^{(i)}(\cdot\mid y_{t-1})$, with respective weights $\pi_{0}(y_{t-1})$ and $\pi_{1}(y_{t-1})=1-\pi_{0}(y_{t-1})$. 
The baseline component is common to all mice from genotype $x_{1}$ with $z_{1,x_{1}}=h_{1}$ singing under context $x_{2}$ with $z_{2,x_{2}}=h_{2}$, 
and provides a type of fixed effects term.  The probability $\pi_1(y_{t-1})$ characterizes the 
amount of heterogeneity among mice, taking the place of the random effects variance in a traditional mixed effects model.  The 
convex structure facilitates computation and interpretability.

\begin{figure}[h!]
\begin{center}
\includegraphics[height=8cm, width=15cm, trim=1cm 1cm 1cm 1cm, clip=true]{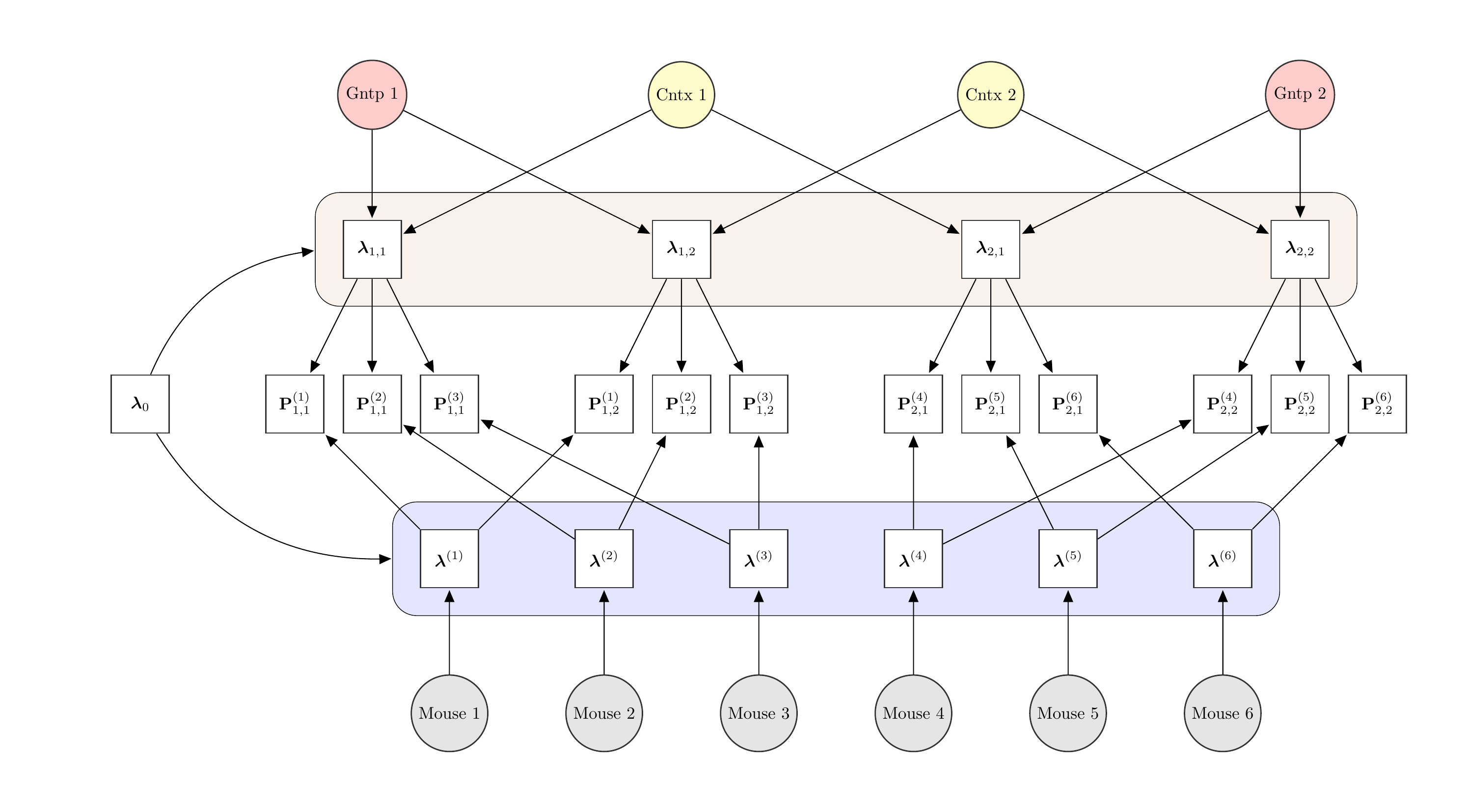}
\end{center}
\vspace{-0.5cm}
\caption{Graphical model showing information sharing for sequences with similar predictors and/or from the same 
mouse.  Predictors include genotype $\in \{\mbox{Gntp 1, Gntp 2} \}$ and context $\in \{ \mbox{Cntx 1, Cntx 2} \}$.
Mice 1-3 have Gntp 1, Mice 4-6 have Gntp 2.  Brown shading = predictor effects, Blue shading = random effects.}
\label{fig: information sharing}
\end{figure}

Figure \ref{fig: information sharing} shows how the proposed formulation shares information across songs associated with the same levels of genotype and context 
as well as across songs sung by the same mouse.  The convex form of the model is interpretable as a two-component mixture of a global and local component. 
Related global-local mixtures have been proposed in fundamentally different contexts by  \cite{muller_etal:2004}, \cite{dunson:2006} and \cite{ren_etal:2010}.

\begin{figure}[h!]
\begin{center}
\includegraphics[height=12cm, width=8.5cm, trim=1cm 1cm 1cm 1cm, clip=true]{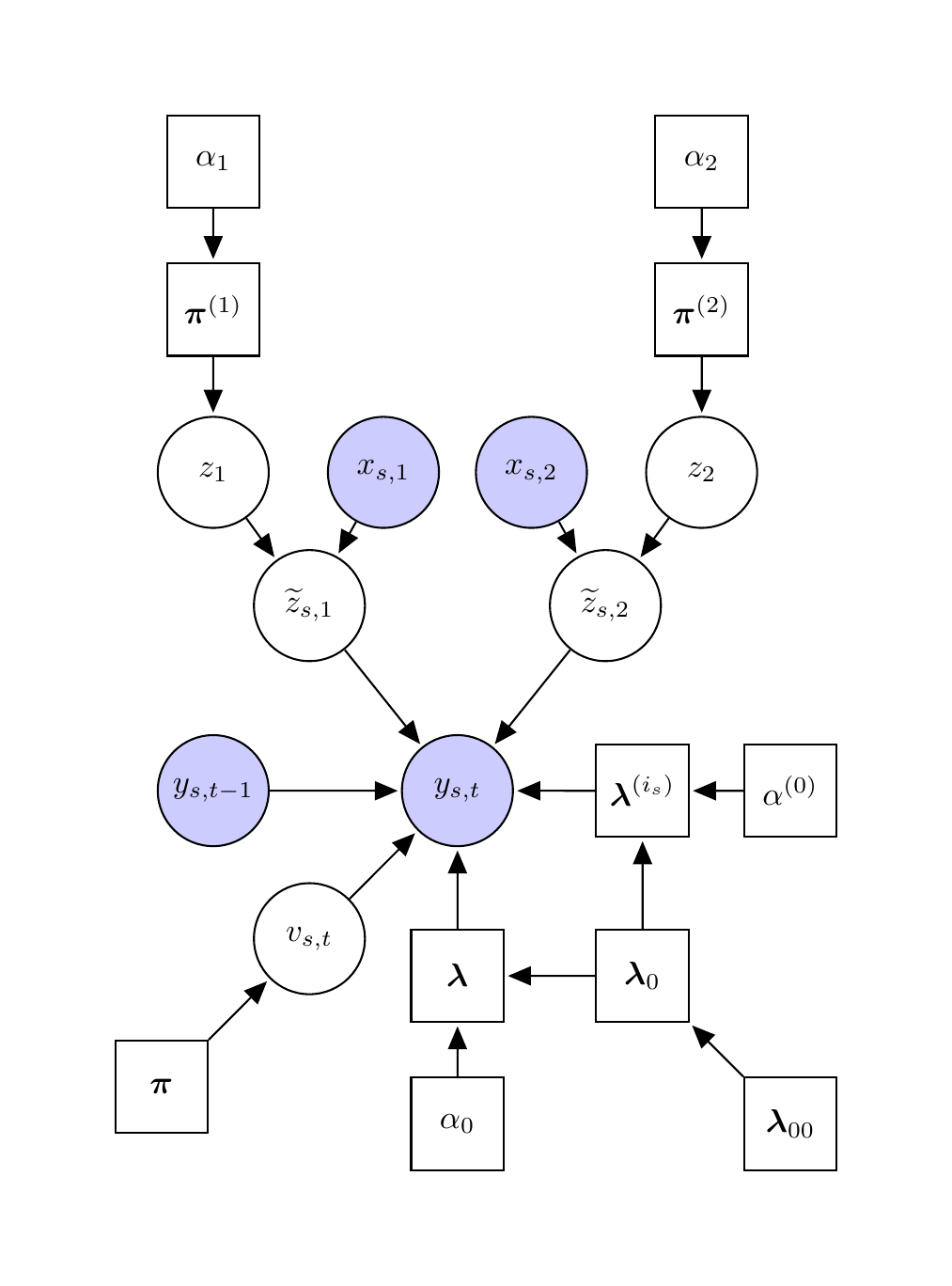}
\end{center}
\vspace{-0.5cm}
\caption{Pictorial representation of the local dependency structure in a sequence $\{y_{s,t}\}$ produced by the $i_{s}\th$ individual with two exogenous predictors $x_{s,1}$ and $x_{s,2}$. 
Observed and latent variables are marked by shaded and unshaded nodes, respectively.
}
\label{fig: local dependency}
\end{figure}

\subsection{Model Properties}\label{sec: properties}

Switching back to general settings with $p$ exogenous predictors, we have
\bse
&& \bP_{h_{1},\dots,h_{p}}^{(i)} (\cdot\mid y_{t-1}) = \pi_{0}(y_{t-1}) \blambda_{h_{1},\dots,h_{p}}(\cdot\mid y_{t-1}) + \pi_{1}(y_{t-1}) \blambda^{(i)}(\cdot\mid y_{t-1}). \label{eq: memcep}
\ese
Introducing latent variables $v_{s,t} \in \{0,1\}$ for each $y_{s,t}$, we can write 
\bse
&& \{y_{s,t} \mid y_{s,t-1}, i_{s}=i, z_{j,x_{s,j}} =h_{j}, j=1,\dots,p, v_{s,t}\}  \sim \nonumber\\
&& \hspace{1cm}  \left\{\begin{array}{lc} 
\Mult[\{1,\dots,d_{0}\},\lambda^{(i)}(1\mid y_{s,t-1}),\dots,\lambda^{(i)}(d_{0}\mid y_{s,t-1})]   & \text{if}~v_{s,t}=0, \\
\Mult[\{1,\dots,d_{0}\},\lambda_{h_{1},h_{2}}(1\mid y_{s,t-1}),\dots,\lambda_{h_{1},h_{2}}(d_{0}\mid y_{s,t-1})]   & \text{if}~v_{s,t}=1, \\
\end{array}\right.\\
&& (v_{s,t}=\ell \mid y_{s,t-1}) \sim \Mult[\{0,1\},\pi_{0}(y_{s,t-1}),\pi_{1}(y_{s,t-1})]. 
\ese
Sampling the $v_{s,t}$'s facilitates posterior computation. A graphical model characterization is shown in  Figure \ref{fig: local dependency}. 

Conditioning on $\blambda_{0}$, an alternative representation of $\bP_{h_{1},\dots,h_{p}}^{(i)}(\cdot \mid y_{t-1})$ is obtained after some simple algebraic manipulation as 
\bse
&& \hspace{-1cm} \bP_{h_{1},\dots,h_{p}}^{(i)}(\cdot\mid y_{t-1}) =  \blambda_{0}(\cdot\mid y_{t-1}) + \pi_{0}(y_{t-1}) \{\blambda_{h_{1},\dots,h_{p}}(\cdot\mid y_{t-1}) - \blambda_{0}(\cdot\mid y_{t-1})\} \nonumber\\
&&~~~~~~~~~~~~~~~~~~~~~~~~~~~~~~~~~~~~~ + \pi_{1}(y_{t-1}) \{\blambda^{(i)}(\cdot\mid y_{t-1}) - \blambda_{0}(\cdot\mid y_{t-1})\}. 
\ese
With $\blambda_{h_{1},\dots,h_{p}}(\cdot\mid y_{t-1})$ and $\blambda^{(i)}(\cdot\mid y_{t-1})$ distributed around $\blambda_{0}(\cdot\mid y_{t-1})$, 
the expressions within the braces above are centered around zero. 
The first term is interpreted as the global mean, the second term as the departure from the global mean due to the fixed effects, 
and the final term as departure due to individual-specific random effects. 
Integrating out $\blambda^{(i)}(\cdot\mid y_{t-1})$, 
the transition probability vectors $\bP_{h_{1},\dots,h_{p}}^{(i)}(\cdot\mid y_{t-1})$ are centered around their population mean 
\bse
&& \bP_{h_{1},\dots,h_{p}}(\cdot\mid y_{t-1}) = \blambda_{0}(\cdot\mid y_{t-1}) + \pi_{0}(y_{t-1}) \{\blambda_{h_{1},\dots,h_{p}} (\cdot\mid y_{t-1})- \blambda_{0}(\cdot\mid y_{t-1})\}.
\ese
Unlike nonlinear GLM based approaches, 
our proposed formulation results in closed form expressions for the population level transition probabilities 
with the fixed effects terms having the same individual and population level interpretations. 
Also, importantly, the population level probability parameters do not depend on the variability of the subject-specific effects. 
These advantages over GLM based approaches are discussed in greater detail in Section S.3 in the Supplementary Materials.

The hyper-parameters $\alpha_{0}$ and $\alpha^{(0)}$ control the variability of $\blambda_{h_{1},\dots,h_{p}}$ and $\blambda^{(i)}$ around their means.
Treating these parameters to be unknown and inferring them from their posterior makes the proposed approach more data adaptive.
For the fixed effects components $\blambda_{h_{1},\dots,h_{p}}$, 
we have 
$$\var \{\lambda_{h_{1},\dots,h_{p}}(y_{t} \mid y_{t-1}) \mid \alpha_{0},\lambda_{0}(y_{t} \mid y_{t-1})\}=\lambda_{0}(y_{t} \mid y_{t-1})\{1-\lambda_{0}(y_{t} \mid y_{t-1})\} (\alpha_{0}+1)^{-1}.$$ 
In the limit, when $\alpha_{0} \to \infty$, $\var \{\lambda_{h_{1},\dots,h_{p}}(y_{t} \mid y_{t-1})  \mid \alpha_{0},\lambda_{0}(y_{t} \mid y_{t-1})\} \to 0$. 
The limiting case $\pi_{0}(y_{t-1}) \to 0$ for all $y_{t-1}\in \S_{0}$ also signifies the absence of fixed effects. 
%
Likewise, for the random effects components, we have 
$$\var \{\lambda^{(i)}(y_{t} \mid y_{t-1}) \mid \alpha^{(0)},\lambda_{0}(y_{t} \mid y_{t-1})\}=\lambda_{0}(y_{t} \mid y_{t-1})\{1-\lambda_{0}(y_{t} \mid y_{t-1})\} (\alpha^{(0)}+1)^{-1}.$$ 
In the limit, when $\alpha^{(0)} \to \infty$, $\var \{\lambda^{(i)}(y_{t} \mid y_{t-1})  \mid \alpha^{(0)},\lambda_{0}(y_{t} \mid y_{t-1})\} \to 0$. 
Random effects are also absent when $\pi_{1}(y_{t-1}) \to 0$ for all $y_{t-1} \in \S_{0}$. 
If interest lies in testing the presence or absence of random effects, 
the prior for $\blambda^{(i)}(\cdot \mid y_{t-1})$ can be adapted to include a point mass at $\blambda_{0}(\cdot \mid y_{t-1})$.

The correlation structure between transition probabilities for different levels of the predictors and individuals 
provides further insights into the information sharing properties of the proposed model. 
For $(h_{11},\dots,h_{1p}) \neq (h_{21},\dots,h_{2p})$, we have 
\bse
\rho^{(i)} &=& \corr\{P_{h_{11},\dots,h_{1p}}^{(i)}(y_{t} \mid y_{t-1}),P_{h_{21},\dots,h_{2p}}^{(i)}(y_{t} \mid y_{t-1})  \mid \lambda_{0}(y_{t} \mid y_{t-1}),\pi_{0}(y_{t-1}),\alpha_{0},\alpha^{(0)}\} \\
&=& \frac{\pi_{1}^{2}(y_{t-1})/(\alpha^{(0)}+1)}   {\{\pi_{0}^{2}(y_{t-1})/(\alpha_{0}+1)+\pi_{1}^{2}(y_{t-1})/(\alpha^{(0)}+1)\} }.
\ese
Likewise, for $i_{1} \neq i_{2}$, we have 
\bse
\rho^{(i_{1},i_{2})} &=& \corr\{P_{h_{1},\dots,h_{p}}^{(i_{1})}(y_{t} \mid y_{t-1}),P_{h_{1},\dots,h_{p}}^{(i_{2})}(y_{t} \mid y_{t-1})  \mid \lambda_{0}(y_{t} \mid y_{t-1}),\pi_{0}(y_{t-1}),\alpha_{0},\alpha^{(0)}\} \\
&=& \frac{\pi_{0}^{2}(y_{t-1})/(\alpha_{0}+1)}   {\{\pi_{0}^{2}(y_{t-1})/(\alpha_{0}+1)+\pi_{1}^{2}(y_{t-1})/(\alpha^{(0)}+1)\} }.
\ese
Finally, for $(h_{11},\dots,h_{1p}) \neq (h_{21},\dots,h_{2p})$ and $i_{1} \neq i_{2}$, we have 
\bse
\corr\{P_{h_{11},\dots,h_{1p}}^{(i_{1})}(y_{t} \mid y_{t-1}),P_{h_{21},\dots,h_{2p}}^{(i_{2})}(y_{t} \mid y_{t-1})  \mid \lambda_{0}(y_{t} \mid y_{t-1}),\pi_{0}(y_{t-1}),\alpha_{0},\alpha^{(0)}\} = 0.
\ese
These expressions are all independent of the mean $\lambda_{0}(y_{t} \mid y_{t-1})$. 
If $\{\pi_{0}^{2}(y_{t-1})(\alpha^{(0)}+1)\}/\{\pi_{1}^{2}(y_{t-1})(\alpha_{0}+1)\} \to 0$, $\rho^{(i)} \to 1$. 
Conversely, if $\{\pi_{0}^{2}(y_{t-1})(\alpha^{(0)}+1)\}/\{\pi_{1}^{2}(y_{t-1})(\alpha_{0}+1)\} \to \infty$, $\rho^{(i)} \to 0$ . 
If $\{\pi_{0}^{2}(y_{t-1})(\alpha^{(0)}+1)\}/\{\pi_{1}^{2}(y_{t-1})(\alpha_{0}+1)\} \to 0$, $\rho^{(i_{1},i_{2})} \to 0$, 
and, conversely, if $\{\pi_{0}^{2}(y_{t-1})(\alpha^{(0)}+1)\}/\{\pi_{1}^{2}(y_{t-1})(\alpha_{0}+1)\} \to \infty$, $\rho^{(i_{1},i_{2})} \to 1$. 
Without the random effect components, we would have had $\rho^{(i)}=0$ and $\rho^{(i_{1},i_{2})}=1$.

Supplementary Materials contain further details on some properties of our proposed model, including ergodicity, flexibility and posterior consistency.  As the model is effectively nonparametric, we can show that the transition probabilities for each of the subjects are estimated consistently as the length of each sequence increases. 
This seems to be the appropriate asymptotic notion in our motivating application, as it is easy to collect more data on mice but practically impossible to study more than a small to moderate number of mice.

\section{Posterior Inference}\label{sec: pi}

\subsection{Posterior Computation}\label{sec: pc}
Inference is based on samples drawn from the posterior using a Gibbs sampler that exploits the conditional independence relationships depicted in Figure \ref{fig: local dependency}. 
In what follows, $\bzeta$ denotes a generic variable that collects all other variables not explicitly mentioned, including the data points.
The sampler comprises the following steps. 

\begin{enumerate}[leftmargin=0cm,itemindent=.5cm,labelwidth=\itemindent,labelsep=0cm,align=left]
\item 
Sample each $z_{j,\ell}$ according to its multinomial full conditional  
\bse
&& p(z_{j,\ell}=h_{j} \mid z_{j',\ell}=h_{j'}, j'\neq j, \bzeta) \propto \pi_{h_{j}}^{(j)}  \times \\
&& \prod_{y_{t-1}}\prod_{(h_{1},\dots,h_{p})}  \frac{\beta\{\alpha_{0}\lambda_{0}(1 \mid y_{t-1})+n_{h_{1},\dots,h_{p}}(1 \mid y_{t-1}), \dots, \alpha_{0}\lambda_{0}(d_{0} \mid y_{t-1})+n_{h_{1},\dots,h_{p}}(d_{0} \mid y_{t-1})\}}{\beta\{\alpha_{0}\lambda_{0}(1 \mid y_{t-1}),\dots,\alpha_{0}\lambda_{0}(d_{0} \mid y_{t-1})\}},
\ese
where $n_{h_{1},\dots,h_{p}}(y_{t} \mid y_{t-1}) = \sum_{s,t}1\{y_{s,t}=y_{t}, y_{s,t-1}=y_{t-1},v_{s,t}=0, z_{j,x_{s,j}}=h_{j}, j=1,\dots,p\}$. 

\item 
Sample each $\bpi^{(j)}$ according to its Dirichlet full conditional  
\bse
&& \{\pi^{(j)}(1),\dots,\pi^{(j)}(d_{j})\} \mid \bzeta \sim \Dir\{\alpha_{j}+n_{j}(1),\dots,\alpha_{j}+n_{j}(d_{j})\},
\ese
where $n_{j} (h)= \sum_{\ell=1}^{d_{j}}1\{z_{j,\ell}=h\}$. 

\item 
Sample each $v_{s,t}$ according to its Bernoulli full conditional  
\bse
&& p(v_{s,t}=v \mid \bzeta) \propto \pi_{v}(y_{s,t-1})  \times \wt\lambda_{v}(y_{s,t} \mid y_{s,t-1}),
\ese
where $\wt\blambda_{0}(\cdot \mid y_{t-1})=\blambda_{h_{1},\dots,h_{p}}(\cdot \mid y_{t-1})$ with $(z_{1,x_{s,1}},\dots,z_{p,x_{s,p}})=(h_{1},\dots,h_{p})$, and $\wt\blambda_{1}(\cdot \mid y_{t-1})=\blambda^{(i)}(\cdot \mid y_{t-1})$.

\item 
Sample $\bpi = \{\pi_{0}(y_{t-1}), \pi_{1}(y_{t-1})\}\trans$ according to its Beta full conditional  
\bse
&& \{\pi_{0}(y_{t-1}),\pi_{1}(y_{t-1})\} \mid \bzeta \sim \Beta\{a_{0}+n_{0}(y_{t-1}),a_{1}+n_{1}(y_{t-1})\},
\ese
where $n_{v} (y_{t-1})= \sum_{s,t}1\{v_{s,t}=v, y_{s,t-1}=y_{t-1}\}$. 

\item 
Sample each $\blambda^{(i)}(\cdot \mid y_{t-1})$'s according to its Dirichlet full conditional 
\bse
&& \hspace{-1cm} \{\lambda^{(i)}(1 \mid y_{t-1}),\dots,\lambda^{(i)}(d_{0} \mid y_{t-1})\} \mid \bzeta \sim \\
&&	\Dir\{\alpha^{(0)} \lambda_{0}(1 \mid y_{t-1})+n^{(i)}(1 \mid y_{t-1}),\dots,\alpha^{(0)} \lambda_{0}(d_{0} \mid y_{t-1})+n^{(i)}(d_{0} \mid y_{t-1})\},
\ese
where $n^{(i)}(y_{t} \mid y_{t-1}) = \sum_{s,t}1\{y_{s,t}=y_{t}, y_{s,t-1}=y_{t-1}, v_{s,t}=1, i_{s}=i\}$.

\item 
Sample each $\blambda_{h_{1},\dots,h_{p}}(\cdot \mid y_{t-1})$ according to its Dirichlet full conditional 
\bse
&& \hspace{-1cm} \{\lambda_{h_{1},\dots,h_{p}}(1 \mid y_{t-1}),\dots,\lambda_{h_{1},\dots,h_{p}}(d_{0} \mid y_{t-1})\} \mid \bzeta \sim \\
&&	\Dir\{\alpha_{0}\lambda_{0}(1 \mid y_{t-1})+n_{h_{1},\dots,h_{p}}(1 \mid y_{t-1}),\dots,\alpha_{0}\lambda_{0}(d_{0} \mid y_{t-1})+n_{h_{1},\dots,h_{p}}(d_{0} \mid y_{t-1})\}.
\ese

\item 
Let $v_{h_{1},\dots,h_{p}}(y_{t} \mid y_{t-1})=0$ and $n=0$. 
Sample an auxiliary variable $\xi$ as
\bse
\xi \mid \bzeta \sim \Bern\left\{\frac{\alpha_{0}\lambda_{0}(y_{t} \mid y_{t-1})}{n+\alpha_{0}\lambda_{0}(y_{t} \mid y_{t-1})}\right\}.
\ese
Set $n=n+1$ and $v_{h_{1},\dots,h_{p}}(y_{t} \mid y_{t-1})=v_{h_{1}\dots,h_{p}}(y_{t} \mid y_{t-1})+\xi$. 
Continue until $n$ equals $n_{h_{1},\dots,h_{p}}(y_{t} \mid y_{t-1})$. 
Likewise, set $v^{(i)}(y_{t} \mid y_{t-1})=0$ and $n=0$. 
Sample an auxiliary variable $\xi$ as
\bse
\xi \mid \bzeta \sim \Bern\left\{\frac{\alpha^{(0)}\lambda_{0}(y_{t} \mid y_{t-1})}{n+\alpha^{(0)}\lambda_{0}(y_{t} \mid y_{t-1})}\right\}.
\ese
Set $n=n+1$ and $v^{(i)}(y_{t} \mid y_{t-1})=v^{(i)}(y_{t} \mid y_{t-1})+\xi$. 
Continue until $n$ equals $n^{(i)}(y_{t} \mid y_{t-1})$. 
Set 
$v(y_{t} \mid y_{t-1}) = \sum_{h_{1},\dots,h_{p}}v_{h_{1},\dots,h_{p}}(y_{t} \mid y_{t-1}) + \sum_{i} v^{(i)}(y_{t} \mid y_{t-1})$.
Also, 
set $v_{0}=\sum_{y_{t}}\sum_{y_{t-1}}\sum_{h_{1},\dots,h_{p}} v_{h_{1},\dots,h_{p}}(y_{t} \mid y_{t-1})$, and 
$v^{(0)}=\sum_{y_{t}}\sum_{y_{t-1}}\sum_{h_{1},\dots,h_{p}}v^{(i)}(y_{t} \mid y_{t-1})$. 
Additionally, sample auxiliary variables 
\bse
r_{h_{1},\dots,h_{p}}(y_{t-1}) \mid \bzeta \sim \Beta\{\alpha_{0}+1,n_{h_{1},\dots,h_{p}}(y_{t-1})\}, \\
s_{h_{1},\dots,h_{p}}(y_{t-1}) \mid \bzeta \sim \Bern\left\{\frac{n_{h_{1},\dots,h_{p}}(y_{t-1})}{n_{h_{1},\dots,h_{p}}(y_{t-1})+\alpha_{0}}\right\}, \\
r^{(i)}(y_{t-1}) \mid \bzeta \sim \Beta\{\alpha^{(0)}+1,n^{(i)}(y_{t-1})\}, \\
s^{(i)}(y_{t-1}) \mid \bzeta \sim \Bern\left\{\frac{n^{(i)}(y_{t-1})}{n^{(i)}(y_{t-1})+\alpha^{(0)}}\right\},
\ese
where $n_{h_{1},\dots,h_{p}}(y_{t-1}) = \sum_{y_{t}} n_{h_{1},\dots,h_{p}}(y_{t} \mid y_{t-1}) $ and $n^{(i)}(y_{t-1})=\sum_{y_{t}}n^{(i)}(y_{t} \mid y_{t-1})$.
Set 
$\log ~ r_{0}=\sum_{y_{t-1}}\sum_{h_{1},\dots,h_{p}} \log~r_{h_{1},\dots,h_{p}}(y_{t-1})$, 
$s_{0}=\sum_{y_{t-1}}\sum_{h_{1},\dots,h_{p}}s_{h_{1},\dots,h_{p}}(y_{t-1})$, 
$\log ~ r^{(0)}=\sum_{y_{t-1}}\sum_{i} \log~r^{(i)}(y_{t-1})$, and 
$s^{(0)}=\sum_{y_{t-1}}\sum_{m}s^{(i)}(y_{t-1})$.

\item 
Sample $\alpha_{0}$ according to its Gamma full conditional 
\bse
&& \alpha_{0} \mid \bzeta \sim \Ga (a_{\alpha_{0}}+v_{0}-s_{0}, b_{\alpha_{0}}-\log~r_{0}). 
\ese

\item 
Sample $\alpha^{(0)}$ according to its Gamma full conditional 
\bse
&& \alpha^{(0)} \mid \bzeta \sim \Ga (a_{\alpha^{(0)}}+v^{(0)}-s^{(0)}, b_{\alpha^{(0)}}- \log~r^{(0)}). 
\ese

\item 
~Finally, sample $\blambda_{0}$ according to its Dirichlet full conditional 
\bse
&& \hspace{-1cm} \{\lambda_{0}(1 \mid y_{t-1}),\dots,\lambda_{0}(d_{0} \mid y_{t-1})\} \mid \bzeta \sim \\
&& \Dir\{\alpha_{00}\lambda_{00}(1)+v(1 \mid y_{t-1}),\dots,\alpha_{00}\lambda_{00}(d_{0})+v(d_{0} \mid y_{t-1})\}. 
\ese
\end{enumerate}
The steps to update the hyper-parameters $\alpha_{0}$, $\alpha^{(0)}$ and the global transition distributions $\blambda_{0}$ were adapted from the auxiliary variable sampler of \cite{west:1992} and \cite{Teh_etal:2006}. 
In all our examples, $5,000$ MCMC iterations with the initial $2,000$ discarded as burn-in and the remaining samples thinned by an interval of $5$ 
produced very stable estimates of the individual and population level parameters of interest. 
MCMC diagnostic checks were not indicative of any convergence or mixing issues. 
Our implementation is fully automated, 
taking in only a single matrix argument - 
concatenated sequences $y_{s,t}$ with the associated values of the exogenous predictors $x_{s,j}$ and the subject labels
repeated $T_{s}$ times for each sequence $s$ and included as additional columns. 
For the Foxp2 data set, this required feeding a $148778 \times 4$ dimensional data matrix to the codes and $5,000$ MCMC iterations required approximately two hours to run on an ordinary laptop. 
\footnote{These codes will be available from the first and the last authors' github repositories 
and also as part of the Supplementary Materials once the paper is accepted for publication.}

\subsection{Prior Hyper-parameters and MCMC Initializations} \label{sec: prior hyper-hyparameters}
In all our examples, real or synthetic, we set $\alpha_{00}=1$ and $\lambda_{00}(y_{t})=\sum_{s,t}1\{y_{s,t}=y_{t}\}/\sum_{s}T_{s}$, the overall proportion of syllables among all songs. 
We set each $\alpha_{j}$ at the value for which $p_{0}(H_{0j})=p_{0}(\wt{k}_{j}=1)=1/2$.  
For $j=1,\dots,p$, we initialize $\bpi^{(j)}$ at $(1/d_{j},\dots,1/d_{j})\trans$. 
We initialize each $z_{j,h}$ at $h$ for $h=1,\dots,d_{j}$. 
Each level of $x_{j}$ thus initially forms its own cluster. 
The associated $\blambda_{h_{1},\dots,h_{p}}(y_{t} \mid y_{t-1})$ are initialized at $\sum_{s,t}1\{y_{s,t}=y_{t},y_{s,t-1}=y_{t-1},x_{s,j}=h_{j},j=1,\dots,p\}/\sum_{s,t}1\{y_{s,t-1}=y_{t-1},x_{s,j}=h_{j},j=1,\dots,p\}$. 
Likewise, $\blambda^{(i)}(y_{t} \mid y_{t-1})$ are initialized at $\sum_{s,t}1\{y_{s,t}=y_{t},y_{s,t-1}=y_{t-1},i_{s}=i\}/\sum_{s,t}1\{y_{s,t-1}=y_{t-1},i_{s}=i\}$. 
For each $y_{t-1}$, $\{\pi_{0}(y_{t-1}),\pi_{1}(y_{t-1})\}$ is initialized at $(0.8,0.2)$. 
The $v_{s,t}$'s are initialized by sampling from Bernoulli distribution with parameter $\pi_{0}(y_{s,t-1})$. 
The parameters $\alpha_{0}$ and $\alpha^{(0)}$ are both initialized at $1$. 
For the remaining fixed hyper-parameters, we set $a=b=a_{\alpha_{0}}=b_{\alpha_{0}}=a_{\alpha^{(0)}}=b_{\alpha^{(0)}}=1$. 
Extensive experiments suggested the results to be highly robust to these choices. 

\subsection{Assessment of Global and Local Differences} \label{sec: testing}
Following Section \ref{sec: pm for mcep}, $\wt{k}_{j}>1$ if and only if $H_{1j}$ is true, so that the posterior probability of 
$H_{1j}$ can be estimated simply as the proportion of Gibbs sampling draws for which $\wt{k}_{j}>1$.  If there is substantial 
evidence in favor of $H_{1j}$, it is typically of additional interest to conduct pairwise comparisons.  
For example, in the Foxp2 data set, if the effects of genotype $x_{1}$ are significant, 
interest lies in assessing the differences between $P_{1,x_{2}}(y_{t} \mid y_{t-1})$ and $P_{2,x_{2}}(y_{t} \mid y_{t-1})$ 
locally for each $(y_{t}, y_{t-1}) \in \S_{0}^{2}$ for each fixed context $x_{2} \in \S_{2}$. 
In particular, we consider the local tests
$H_{0,y_{t} \mid y_{t-1},x_{2}} : \abs{\Delta P_{\cdot,x_{2}}(y_{t} \mid y_{t-1})}=\vert P_{1,x_{2}}(y_{t} \mid y_{t-1})-P_{2,x_{2}}(y_{t} \mid y_{t-1}) \vert \le \delta$ vs 
$H_{1,y_{t} \mid y_{t-1},x_{2}} : \abs{\Delta P_{\cdot,x_{2}}(y_{t} \mid y_{t-1})} > \delta$
for different $(y_{t}, y_{t-1}) \in \S_{0}^{2}$ and $x_{2} \in \S_{2}$, with $\delta$ a small constant elicited to correspond to a 
`practically' significant difference.  The posterior probabilities of these local tests can be easily estimated from the output of
the Gibbs sampler.

\section{Application to the Foxp2 Data Set} \label{sec: afds}

In this section, we discuss the results of the proposed methodology applied to the Foxp2 data set.  Our inference 
goals focus on studying the systematic variation in the mouse song dynamics across genotypes and contexts, with it being 
of additional substantial interest to assess the magnitude of unexplained variation among mice.

\begin{figure}[h!]
\begin{center}
\includegraphics[height=8.6cm, width=15cm, trim=4cm 1cm 3cm 1cm, clip=true]{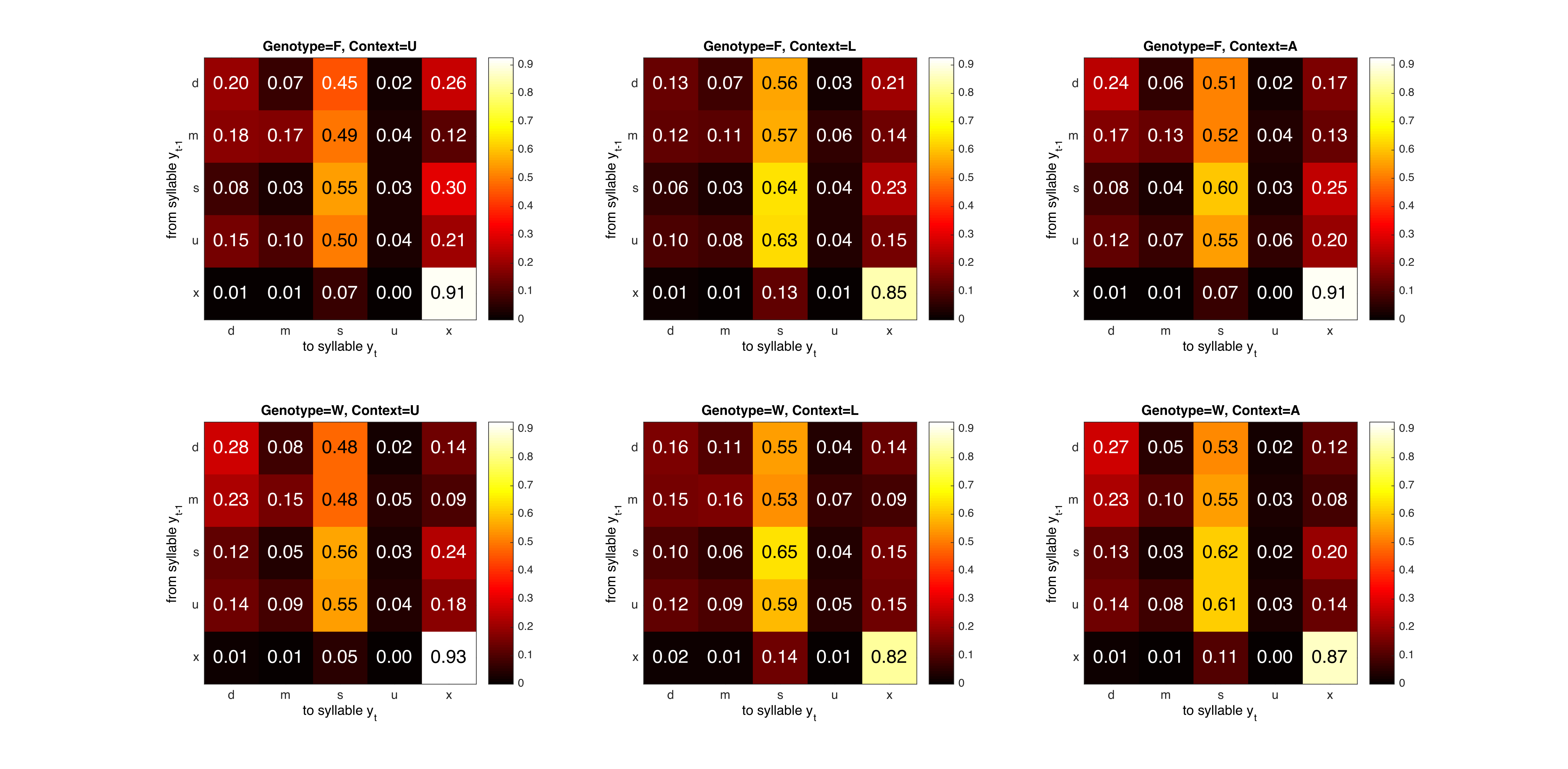}
\caption{Results for the Foxp2 data set. 
Estimated posterior mean transition probabilities $P_{x_{1},x_{2}}(y_{t} \mid y_{t-1})$ for syllables $y_{t},y_{t-1} \in \{d,m,s,u,x\}$ for different combinations of genotype $x_{1} \in \{F,W\}$ and social contexts $x_{2} \in \{U,L,A\}$.}
\label{fig: Foxp2 post mean}
\vskip 15pt
\includegraphics[height=4cm, width=15cm, trim=4cm 0cm 3cm 0cm, clip=true]{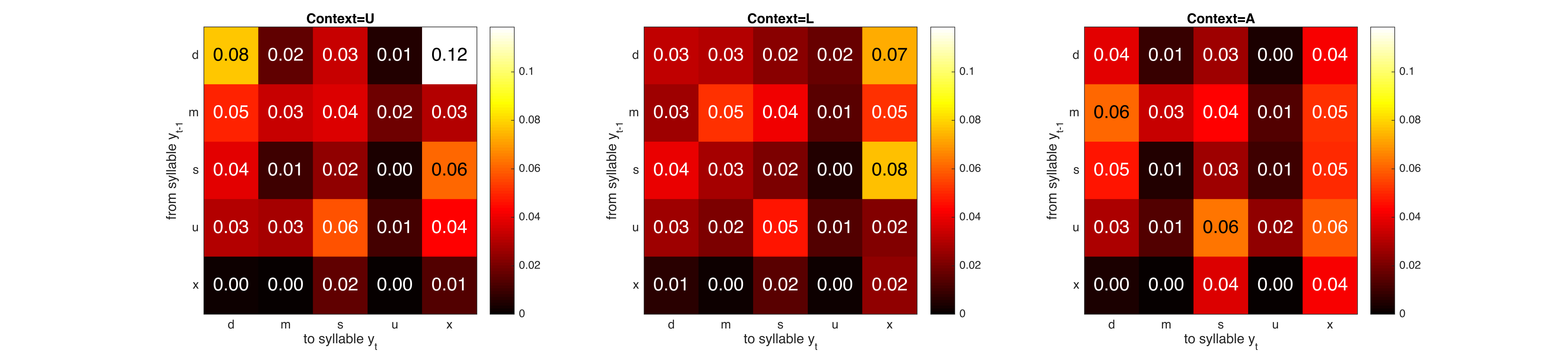}
\vspace*{0.5cm}
\includegraphics[height=4cm, width=15cm, trim=4cm 0cm 3cm 0cm, clip=true]{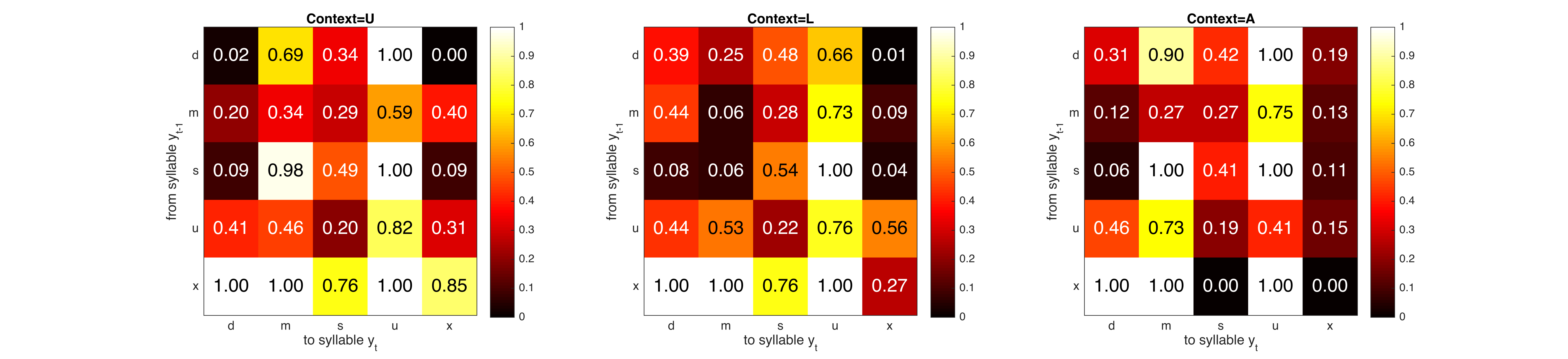}
\vspace*{-0.5cm}
\caption{Results for the Foxp2 data set. 
The top row shows the estimated posterior mean of $\abs{\Delta P_{\cdot,x_{2}}(y_{t} \mid y_{t-1})}=\abs{P_{1,x_{2}}(y_{t} \mid y_{t-1}) - P_{2,x_{2}}(y_{t} \mid y_{t-1})}$ 
for syllables $y_{t},y_{t-1} \in \{d,m,s,u,x\}$ and different social contexts $x_{2} \in \{U,L,A\}$.
The bottom row shows the estimated posterior probability of $H_{0,y_{t} \mid y_{t-1},x_{2}} : \abs{\Delta P_{\cdot,x_{2}}(y_{t} \mid y_{t-1})} \leq 0.02$.}
\label{fig: Foxp2 post prob}
\end{center}
\end{figure}

Figure \ref{fig: Foxp2 post mean} shows the estimated posterior mean transition probabilities $P_{x_{1},x_{2}}(y_{t} \mid y_{t-1})$, 
Figures S.1 and S.3 in the Supplementary Materials 
summarize the posterior standard deviations of $P_{x_{1},x_{2}}(y_{t} \mid y_{t-1})$ and the random effects parameters $\pi_{1}(y_{t-1})\lambda^{(i)}(y_{t} \mid y_{t-1})$, respectively. 
The results showed very strong evidence of an effect of both genotype and context on the song dynamics, with 
$\widehat{P}(H_{1j} \mid \Data) \approx 1$ for $j=1,2$.  Hence, there is clear evidence in the data that knocking out the Foxp2 gene
impacts the mouse vocalization dynamics across the different experimental contexts, with there also being clear evidence that context 
plays a significant role.

We next focus on assessing how the absence of the Foxp2 gene influences the transition probabilities of each of the different transition types locally for each fixed context.  
Figure \ref{fig: Foxp2 post prob} summarizes the posterior mean absolute differences $\abs{\Delta P_{\cdot,x_{2}}(y_{t} \mid y_{t-1})}$ 
and the posterior probabilities of the corresponding local null hypotheses 
$H_{0,y_{t} \mid y_{t-1},x_{2}} : \abs{\Delta P_{\cdot,x_{2}}(y_{t} \mid y_{t-1})} \leq \delta$. 
For the Foxp2 experiment, $\abs{\Delta P_{\cdot,x_{2}}(y_{t} \mid y_{t-1})}$ not exceeding $\delta=0.02$ was assumed to be practically insignificant. 
Results were robust to the choice of this difference threshold. 

The transition types for which the corresponding local null hypotheses $H_{0,\ell}$ had estimated posterior probabilities smaller than $0.10$ were 
$d \to d, s \to d, d \to x, s \to x$ for context $U$; 
$s \to d, m \to m, s \to m, d \to x, m \to x, s \to x$ for context $F$; 
and  $s \to d, x \to s, x \to x$ for context $A$. 
Comparisons between the estimated transition probabilities for the two genotypes summarized in Figure \ref{fig: Foxp2 post mean} 
suggest that, except for transitions from silence to silence in the $U$ context, 
the estimated transition probabilities $P_{x_{1},x_{2}}(x \mid y_{t-1})$ of moving to the special silence syllable `$x$' from a preceding syllable $y_{t-1}$ 
are consistently larger in the Foxp2 mutant mice ($x_{1}=F$) compared to the wild type mice ($x_{1}=W$) 
for all $y_{t-1} \in \{d,m,s,u,x\}$ across all contexts $x_{2}\in\{U,F,A\}$. 
The local nulls $H_{0,\ell}$ corresponding to six of these transition types had estimated posterior probabilities less than $0.10$, 
and four more in the $A$ context had estimated posterior probabilities less than $0.20$. 
Compared to wild type mice, Foxp2 mutant mice thus had a greater tendency to transition to silence across all contexts.

The increase in the probabilities of moving to the `$x$' syllable in the Foxp2 mutant mice seems to be explained by 
an associated decrease in the probabilities of transitioning to the `$d$' syllable. 
Four of the associated local nulls $H_{0,\ell}$ had estimated posterior probabilities less than $0.10$, 
two more had estimated posterior probabilities less than $0.20$.
Additionally, in the $F$ context, the estimated transition probabilities $P_{x_{1},x_{2}}(m \mid y_{t-1})$ of moving to the most complex syllable `$m$' from a preceding syllable $y_{t-1}$  
were smaller in the Foxp2 mutant mice compared to the wild type mice  
for all $y_{t-1} \in \{d,m,s,u\}$. 
Two of the associated local nulls $H_{0,\ell}$ were significant. 
Overall, these findings are consistent with the hypothesis that the Foxp2 gene plays an important role in the complexity and richness
of the song dynamics, with Foxp2 knockout mice having impaired ability to produce more complex syntax. 

A GLM based approach to mixed effects Markov chains, 
implemented via the MCMCglmm package in R \citep{hadfield:2010}, was also applied to the Foxp2 data set. 
Due to methodological limitations and computational complexities,
we could only perform approximate inference with a restrictive parametric model without any interaction terms.
Global significance of the exogenous predictors could not be straightforwardly assessed by such models. 
The approach developed in Section \ref{sec: testing} to assess local differences in transition probabilities between the two genotypes 
could also be used for the GLM based model.  
However, 
no local difference was found to be significant. 
Details are deferred to Section S.3 in the Supplementary Materials.

\section{Simulation Experiments} \label{sec: se}
We designed simulation experiments to evaluate the performance of the proposed methodology in assessing various aspects of the vocalization dynamics in a wide range of scenarios. 
We tried to closely mimic many aspects of the Foxp2 data set to create scenarios typical in mouse vocalization experiments. 
We chose $d_{0}=5$ syllables $\{d,m,s,u,x\}$, $d_{1}=2$ genotypes $\{F,W\}$ with 8 mice from the first genotype and 6 from the other, 
and $d_{2}=3$ contexts $\{U,L,A\}$.  The sequences were chosen to be of the same length as those in the Foxp2 data set. 
In each case, we set $\bpi_{0}$ and $\blambda^{(i)}$ equal to the corresponding estimated values from the Foxp2 data set. 
To evaluate performance in assessing predictor effects, we considered the following scenarios:
\begin{enumerate}[label=(\Alph*)]
\item 
$\blambda_{x_{1},x_{2}}=\wh\blambda_{1,1}$ for $x_{1}=1,2$ and $x_{2}=1,2,3$.
The vocalization dynamics vary neither with genotype nor with context ($\wt{k}_{10}=1$ and $\wt{k}_{20}=1$).  
\item 
$\blambda_{x_{1},x_{2}}=\wh\blambda_{x_{1},1}$ for $x_{1}=1,2$ and $x_{2}=1,2,3$.
The vocalization dynamics vary with genotype but not with context ($\wt{k}_{10}=2$ and $\wt{k}_{20}=1$). 
\item 
$\blambda_{x_{1},x_{2}}=\wh\blambda_{1,x_{2}}$ for $x_{1}=1,2$ and $x_{2}=1,2,3$.
The vocalization dynamics do not vary with genotype but vary with context ($\wt{k}_{10}=1$ and $\wt{k}_{20}=3$),  
\item 
$\blambda_{x_{1},x_{2}}=\wh\blambda_{x_{1},x_{2}}$ for $x_{1}=1,2$ and $x_{2}=1,2,3$.
The vocalization dynamics vary with both genotype and context ($\wt{k}_{10}=2$ and $\wt{k}_{20}=3$). 
\item 
$\blambda_{x_{1},x_{2}}=\wh\blambda_{x_{1},x_{2}}$ for $x_{1}=1,2$ and $x_{2}=1,2$.
Also, $\blambda_{x_{1},3}=\wh\blambda_{x_{1},2}$ for $x_{1}=1,2$. 
The vocalization dynamics vary with both genotype and context but the dynamics within the contexts 2 and 3 are similar ($\wt{k}_{10}=2$ and $\wt{k}_{20}=2$), 
\item 
$\blambda_{1,x_{2}}=\wh\blambda_{1,x_{2}}$ and $\blambda_{2,x_{2}}=\wh\blambda_{1,x_{2}}+\bDelta_{.,x_{2}}$ for $x_{2}=1,2,3$, 
where many cells in $\bDelta_{.,x_{2}}$ are precisely zero. 
The vocalization dynamics vary with both genotype and context ($\wt{k}_{10}=2$ and $\wt{k}_{20}=3$)
but the differences between the dynamics for the two genotypes are strictly localized in a few specific transition types. 
\end{enumerate}

\begin{figure}[h!]
\begin{center}
\includegraphics[height=7cm, width=14cm, trim=1cm 1cm 1cm 0cm, clip=true]{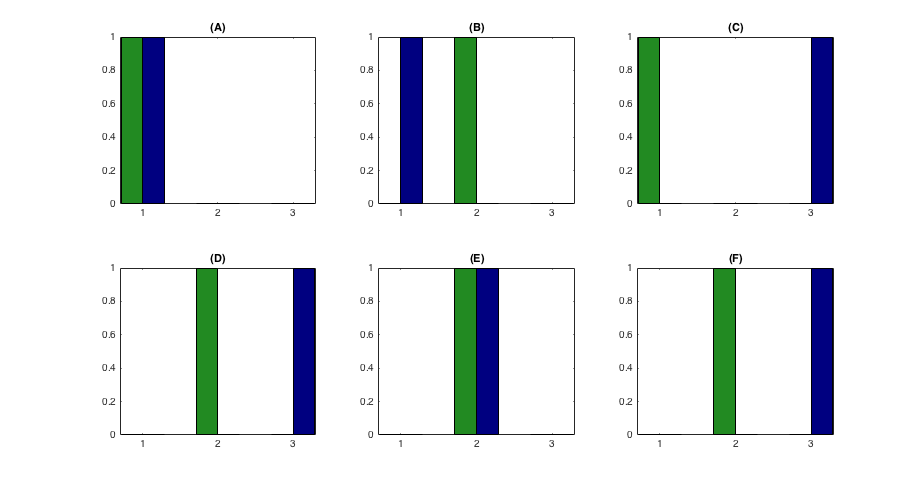}
\end{center}
\vspace{-0.5cm}
\caption{Results of simulation experiments. Estimated posterior probability 
$P(\wt{k}_{j} \mid \Data)=P(\text{the}~d_{j}~\text{levels of}~x_{j}~\text{form}~\wt{k}_{j}~\text{clusters} \mid \Data)$ 
for genotype $x_{1}$ in green and context $x_{2}$ in blue for different simulation scenarios discussed in Section \ref{sec: se}.}
\label{fig: sim study var selection}
\end{figure}

Figure \ref{fig: sim study var selection} shows estimated posterior probabilities 
$\wh{P}(\wt{k}_{j} \mid \Data)=\wh{P}(\wt{k}_j~\text{clusters in}~x_{ij} \mid \Data)$.
For each simulation scenario and $j=1,2$, $\wh{P}(\wt{k}_{j}=\wt{k}_{j0}\mid \Data) \approx 1$.
This excellent performance in global hypothesis testing is not surprising since many small local 
effects can collectively produce strong evidence of overall effects.  

\begin{figure}[hp]
\begin{center}
\includegraphics[height=8.5cm, width=15cm, trim=4cm 1cm 3cm 1cm, clip=true]{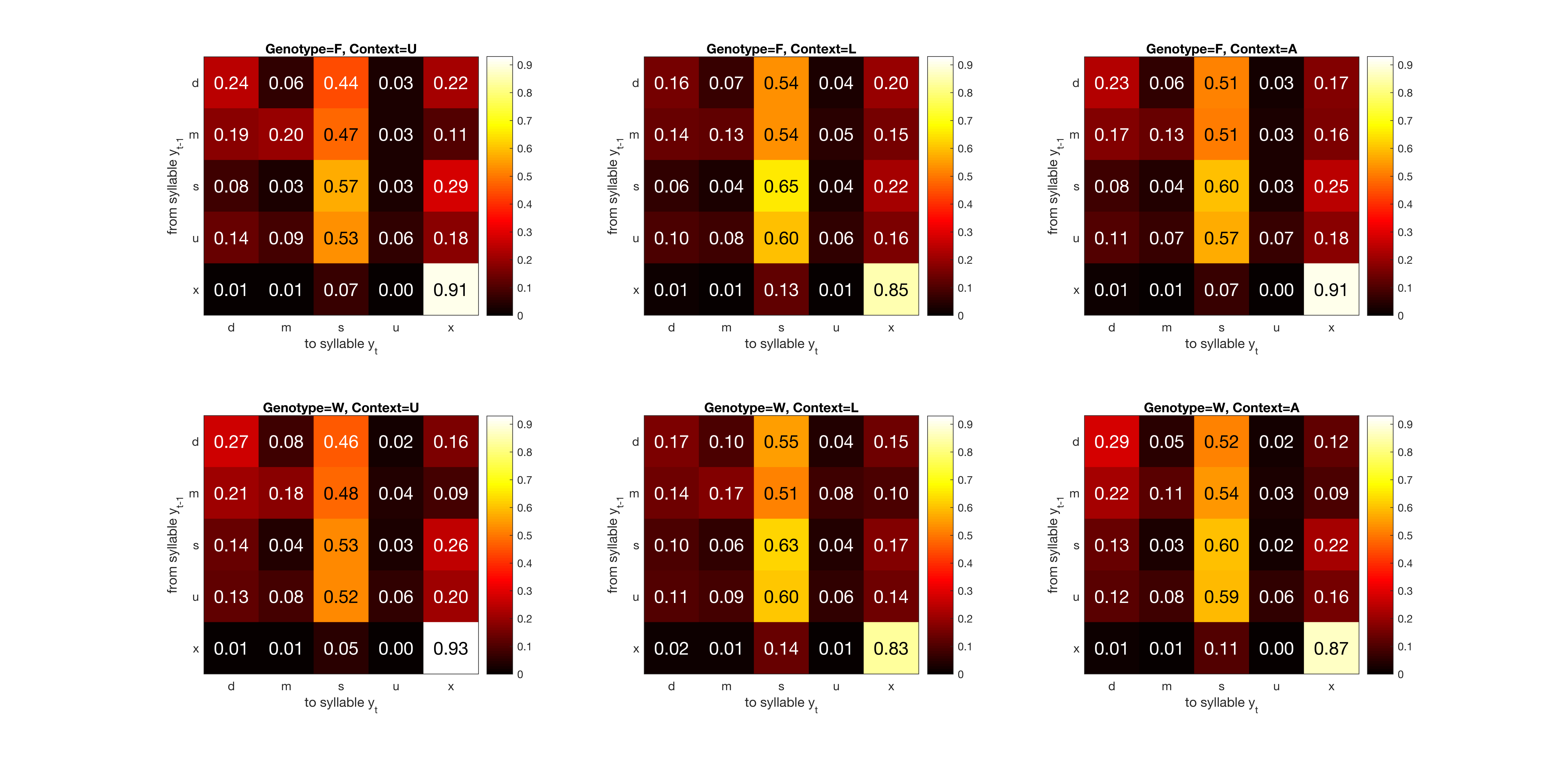}
\caption{Results for the simulation scenario D described in Section \ref{sec: se}. 
Estimated posterior mean transition probabilities $P_{x_{1},x_{2}}(y_{t} \mid y_{t-1})$ for syllables $y_{t},y_{t-1} \in \{d,m,s,u,x\}$ for different combinations of genotype $x_{1} \in \{F,W\}$ and social contexts $x_{2} \in \{U,L,A\}$.}
\label{fig: sim study D post mean}
\vskip 15pt
\includegraphics[height=4cm, width=15cm, trim=4cm 0cm 3cm 0cm, clip=true]{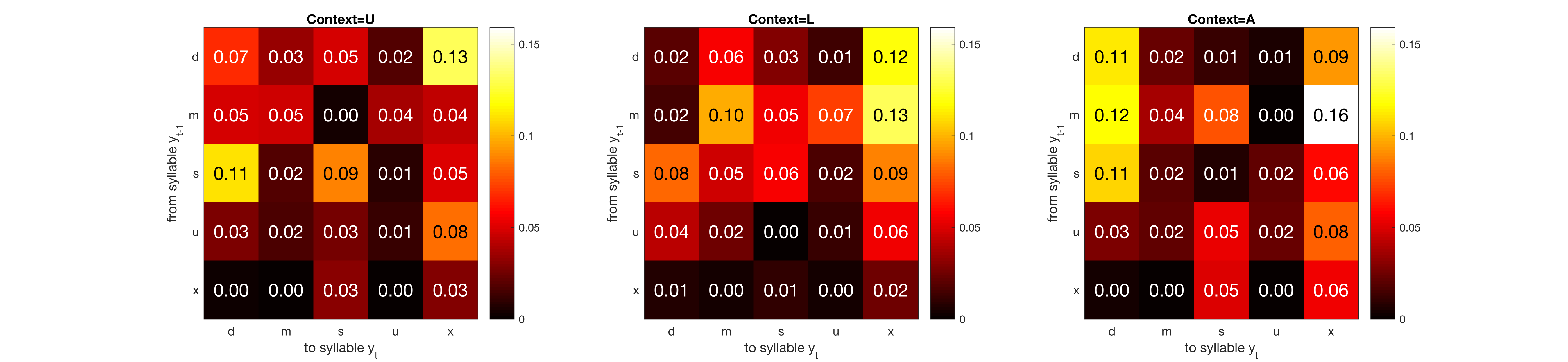}
\vspace*{0.5cm}
\includegraphics[height=4cm, width=15cm, trim=4cm 0cm 3cm 0cm, clip=true]{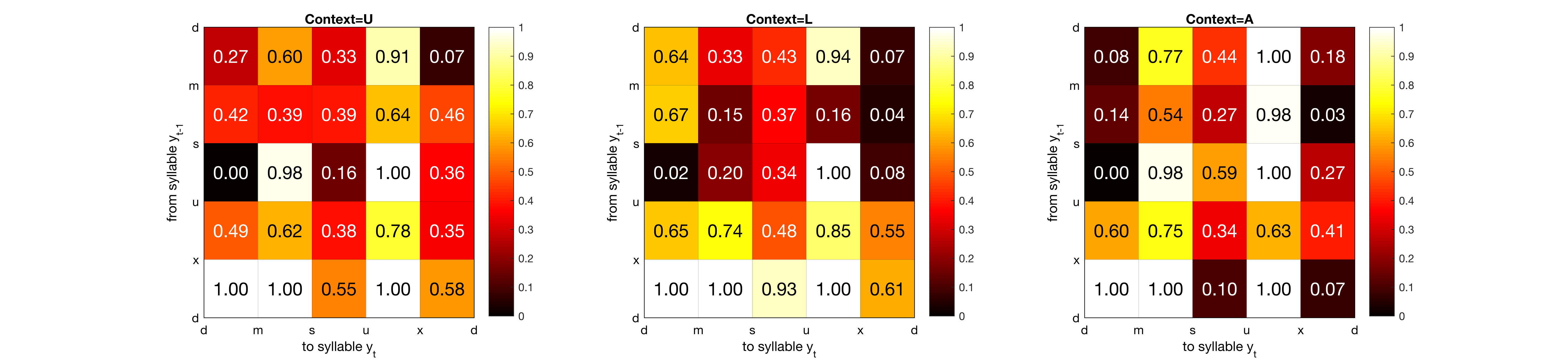}
\caption{Results for the simulation scenario D described in Section \ref{sec: se}. 
The top row shows the estimated posterior mean of $\abs{\Delta P_{\cdot,x_{2}}(y_{t} \mid y_{t-1})}=\abs{P_{1,x_{2}}(y_{t} \mid y_{t-1}) - P_{2,x_{2}}(y_{t} \mid y_{t-1})}$ 
for syllables $y_{t},y_{t-1} \in \{d,m,s,u,x\}$ and social contexts $x_{2} \in \{U,L,A\}$. 
The bottom row shows the estimated posterior probability of $H_{0,y_{t} \mid y_{t-1},x_{2}} : \abs{\Delta P_{\cdot,x_{2}}(y_{t} \mid y_{t-1})} \leq 0.02$.}
\label{fig: sim study D post prob}
\end{center}
\end{figure}

Scenario D was closest to the Foxp2 data.  
Figure \ref{fig: sim study D post mean} shows estimated posterior mean transition probabilities, 
Figures S.2 and S.4 in the Supplementary Materials show estimated posterior standard deviations 
of these probabilities and the random effects parameters, respectively. 
Figure \ref{fig: sim study D post prob} shows estimated posterior mean absolute differences 
$\abs{\Delta P_{\cdot,x_{2}}(y_{t} \mid y_{t-1})}$ 
and posterior probabilities of the local nulls $H_{0,y_{t} \mid y_{t-1},x_{2}} : \abs{\Delta P_{\cdot,x_{2}}(y_{t} \mid y_{t-1})} \leq 0.02$. 
Comparisons with Figure \ref{fig: Foxp2 post mean}, Figure \ref{fig: Foxp2 post sds}, Figure \ref{fig: Foxp2 post sds random effects} and Figure \ref{fig: Foxp2 post prob} 
show remarkable agreement between the results, suggesting high stability and reproducibility.  

\begin{figure}[hp]
\begin{center}
\includegraphics[height=4.5cm, width=15cm, trim=4cm 0cm 3cm 0cm, clip=true]{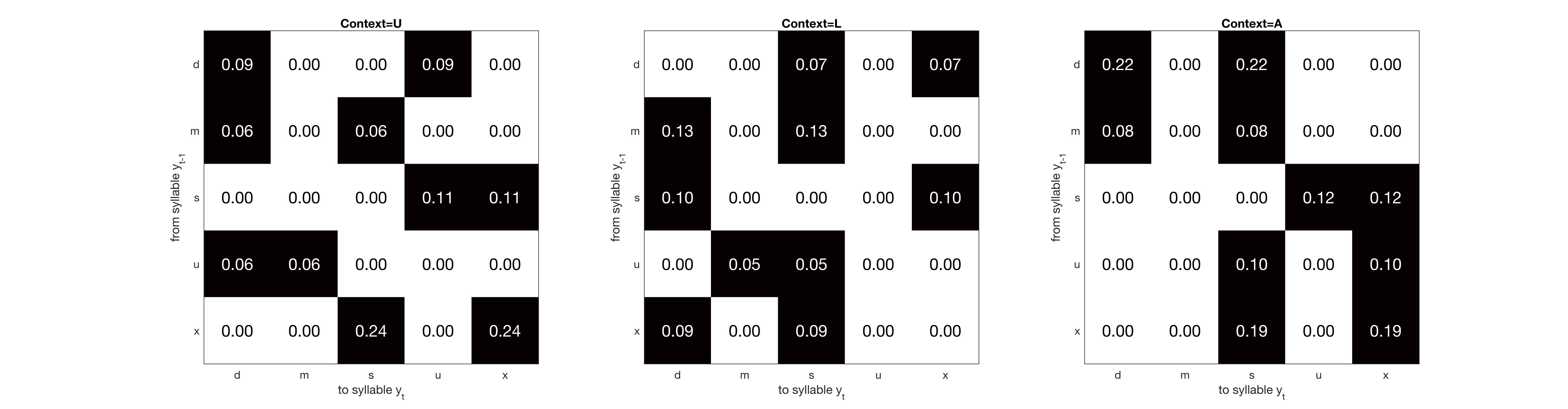}
\includegraphics[height=4.5cm, width=15cm, trim=4cm 0cm 3cm 0cm, clip=true]{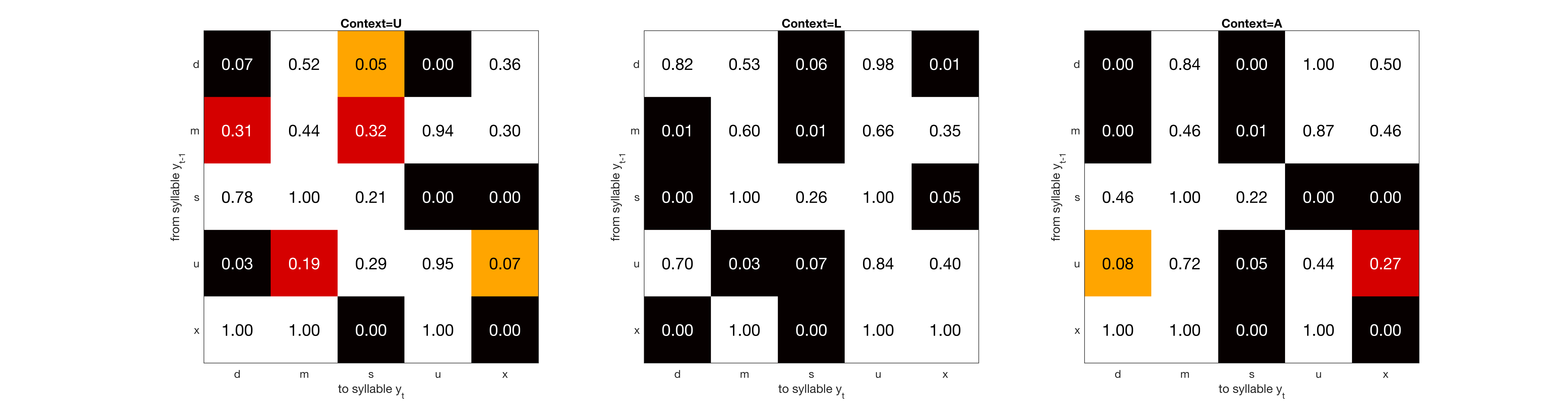}
\includegraphics[height=4.5cm, width=15cm, trim=4cm 0cm 3cm 0cm, clip=true]{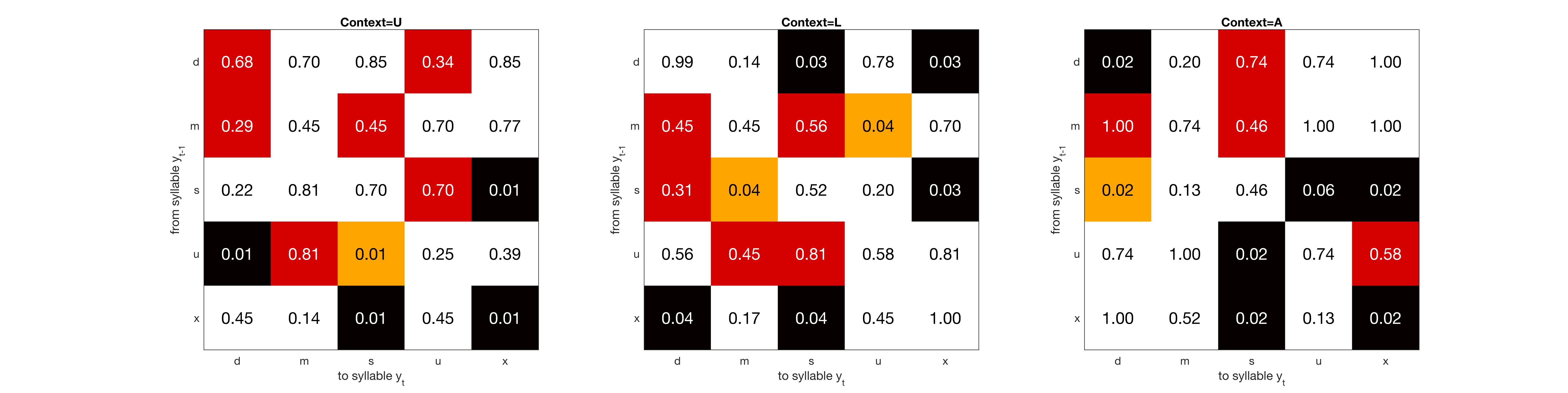}
\caption{Results for the simulation scenario F described in Section \ref{sec: se}. 
The top row shows the true values of $\abs{\Delta P_{\cdot,x_{2}}(y_{t} \mid y_{t-1})}=\abs{P_{1,x_{2}}(y_{t} \mid y_{t-1}) - P_{2,x_{2}}(y_{t} \mid y_{t-1})}$ 
for syllables $y_{t},y_{t-1} \in \{d,m,s,u,x\}$ and social contexts $x_{2} \in \{U,L,A\}$. 
Positive differences are highlighted in black. 
The middle row shows the estimated posterior probabilities of $H_{0,y_{t} \mid y_{t-1},x_{2}} : \abs{\Delta P_{\cdot,x_{2}}(y_{t} \mid y_{t-1})} \leq 0.02$. 
The bottom row shows Benjamini-Hochberg adjusted p-values obtained using the method of \cite{Chabout_etal:2016}. 
Posterior probabilities smaller than $0.1$ are considered significant and are highlighted in black and orange.
Posterior probabilities greater than $0.1$ are presented in white and red.
Likewise, p-values smaller than $0.1$ are considered significant and are highlighted in black and orange.
P-values greater than $0.1$ are presented in white and red.
White and black cells represent correct decisions, 
orange cells mark rejections of true $H_{0,\ell}$ (false positives), 
and red cells mark failures to reject false $H_{0,\ell}$ (false negatives). 
}
\label{fig: sim study F post prob}
\end{center}
\end{figure}

Scenario F was designed to evaluate the performance of the proposed method 
in assessing local differences between transition probabilities for the two genotypes for each fixed context. 
Figure \ref{fig: sim study F post prob} shows the true absolute differences 
$\abs{\Delta P_{\cdot,x_{2}}(y_{t} \mid y_{t-1})}$ 
and the posterior probabilities of the local nulls $H_{0,y_{t} \mid y_{t-1},x_{2}} : \abs{\Delta P_{\cdot,x_{2}}(y_{t} \mid y_{t-1})} \leq 0.02$. 
Out of a total of $3 \times 25=75$ local tests, 
correct inferences were obtained in $68$ tests. 
There were, however, $3$ rejections of true $H_{0,\ell}$'s (false positives), 
and $4$ instances of failures to reject false $H_{0,\ell}$'s (false negatives). 

Previously available methods for mouse song syntax analysis include \cite{Chabout_etal:2015,Chabout_etal:2016}.
Using independent Markov models for each song, 
\cite{Chabout_etal:2015} developed chi-squared tests for assessing global differences in syntax between social contexts. 
\cite{Chabout_etal:2016} developed a more advanced summary statistics based approach for assessing global and local syntax differences. 
They first estimated the transition probabilities separately for each sequence. 
For comparisons between genotypes for fixed contexts, they then applied Wilcoxon-Mann-Whitney rank sum tests to these estimates. 
Finally, these local tests were combined using a permutation based Monte Carlo procedure to assess the global influence of genotype. 
Similar strategies were adopted for assessing the influences of social contexts within and across genotypes. 
Uncertainty in estimating the transition probabilities were, however, completely ignored. 
The assumption of independently distributed sequences was also highly unrealistic. 
More realistically, information should be shared across songs 
associated with same values of the exogenous predictors as well as across songs sung by the same mouse, as in our proposed formulation. 

The bottom row in Figure \ref{fig: sim study F post prob} shows Benjamini-Hochberg adjusted local p-values (adjusted for 25 tests for each context) returned by the approach of \cite{Chabout_etal:2016}. 
With a p-value threshold of $0.10$, 
correct inferences were obtained in $56$ of the $75$ local tests. 
There were $4$ rejections of true $H_{0,\ell}$'s (false positives), 
and $15$ failures to reject false $H_{0,\ell}$'s (false negatives). 
Ideally, to compare their p-value based approach with our posterior probability based procedure, 
we should first calibrate the p-values \citep{selke_etal:2001}. 
Since, for comparable evidence levels, posterior probabilities of null hypotheses typically correspond to calibrated p-values smaller by many orders of magnitude \citep{berger_selke:1987},  such calibration significantly reduces the power of these local tests, further deteriorating the results. 
When we used such calibration, even with our extremely liberal p-value threshold of 0.10, no $H_{0,\ell}$ was rejected. 

Results for another simulated data set with similar performances are summarized in Figure S.5 in the Supplementary Materials.  
Comparisons with a multinomial logit based approach are also discussed in Section S.3 of the Supplementary Materials.

\section{Discussion} \label{sec: dscsn}
This article introduced a flexible, computationally efficient mixed effects Markov model, providing a sophisticated framework for inference in mouse vocalization experiments.  
While the focus was on a particular study assessing the effects of a Foxp2 mutation, the methodology is applicable generally to other vocalization experiments having similar data structures and should have significant impact in such settings.  As it is sometimes of interest to assess changes during development \citep{Castellucci_etal:2016,Scattoni_etal_b:2008}, it will be important in ongoing research to generalize the model to allow dynamic changes with age. 
Although our motivation here was mouse vocalization experiments, the methods have potential far beyond such applications.  Natural extensions, which are conceptually straightforward, include accommodation of hidden Markov Models, continuous predictors, and time-varying predictors.

\baselineskip=17pt
\section*{Supplementary Materials}
Supplementary Materials present theoretical properties, and additional figures summarizing the results for the real and the simulated data sets described in Section \ref{sec: afds} and Section \ref{sec: se}. Supplementary Materials also present additional comparisons of the proposed approach with generalized linear mixed model based approaches. 

\section*{Acknowledgments}
This research was partially funded by Grant N00141410245 of the Office of Naval Research.


\newpage
\baselineskip=14pt
\bibliographystyle{natbib}
\bibliography{NHMM,Neuroscience}

\clearpage\pagebreak\newpage
\pagestyle{fancy}
\fancyhf{}
\rhead{\bfseries\thepage}
\lhead{\bfseries SUPPLEMENTARY MATERIALS}

\baselineskip 20pt
\vspace{-0.5cm}
\begin{center}
{\LARGE{Supplementary Materials} 
for\\ {\bf Bayesian Semiparametric Mixed Effects Markov Chains}}
\end{center}

\setcounter{equation}{0}
\setcounter{page}{1}
\setcounter{table}{1}
\setcounter{figure}{0}
\setcounter{section}{0}
\numberwithin{table}{section}
\renewcommand{\theequation}{S.\arabic{equation}}
\renewcommand{\thesubsection}{S.\arabic{section}.\arabic{subsection}}
\renewcommand{\thesection}{S.\arabic{section}}
\renewcommand{\thepage}{S.\arabic{page}}
\renewcommand{\thetable}{S.\arabic{table}}
\renewcommand{\thefigure}{S.\arabic{figure}}
\baselineskip=12pt

\begin{center}
Abhra Sarkar\\
Department of Statistical Science, Duke University, Box 90251, Durham NC 27708-0251, USA\\
abhra.sarkar@duke.edu\\ 
\vskip 10pt 
Jonathan Chabout\\
Department of Neurobiology, Duke University, Durham, NC 27710, USA\\
jchabout.pro@gmail.com\\
\vskip 10pt 
Joshua Jones Macopson\\
Department of Neurobiology, Duke University, Durham, NC 27710, USA\\
joshua.jones.macopson@duke.edu\\
\vskip 10pt 
Erich D. Jarvis\\
Department of Neurobiology, Duke University, Durham, NC 27710, USA\\
Howard Hughes Medical Institute, Chevy Chase, MD 20815, USA\\
The Rockefeller University, New York, NY 10065, USA\\
jarvis@neuro.duke.edu\\
\vskip 10pt 
David B. Dunson\\
Department of Statistical Science, Duke University, Box 90251, Durham NC 27708-0251, USA\\
dunson@duke.edu\\
\end{center}

\baselineskip=15pt

\vskip 10pt 
The Supplementary Materials discuss some theoretical aspects of our model; 
present additional figures summarizing the results for the real and the simulated data sets described in Section \ref{sec: afds} and Section \ref{sec: se} in the main paper; 
and discuss the contrasting features of generalized linear mixed model based approaches with our proposed approach, 
highlighting the latter's many advantages over the former.

\newpage
\section{Theoretical Properties}

In this section, we discuss some theoretical aspects of our proposed model. 
We follow the notations and definitions of the main paper. 


For Markov chains with exogenous predictors (MCEPs), 
the values of the exogenous predictors remain fixed for the entire lengths of the sequences. 
The notions of ergodicity, stationarity etc for predictor free ordinary Markov chains thus extend naturally to MCEPs. 
Let $\calP_{0}=\{P_{0,x_{1},\dots,x_{p}}^{(i)}(y_{t} \mid y_{t-1}): P_{0,x_{1},\dots,x_{p}}^{(i)}(y_{t} \mid y_{t-1})=\pi_{0,0}(y_{t-1})\blambda_{0,x_{1},\dots,x_{p}}(y_{t} \mid y_{t-1})+\pi_{0,1}(y_{t-1})\blambda_{0}^{(i)}(y_{t} \mid y_{t-1})\} \subset \calP$ 
denote the class of transition probability distributions that admit representations similar to our proposed formulation. 
It is straightforward to check that any $\bP_{0,x_{1},\dots,x_{p}}^{(i)}(\cdot \mid y_{t-1}) \in \calP_{0}$ will be ergodic if at least one of the component transition distributions is also so and the associated mixture probability is strictly positive. 
%
%
In particular, if $\blambda_{0,x_{1},\dots,x_{p}}(\cdot \mid y_{t-1})$ and $\blambda_{0}^{(i)}(\cdot \mid y_{t-1})$ are both ergodic 
with stationary distributions 
$\bpi_{0,x_{1},\dots,x_{p}}=\{\pi_{0,x_{1},\dots,x_{p}}(1),\dots,\pi_{0,x_{1},\dots,x_{p}}(d_{0})\}\trans$ and 
$\bpi_{0}^{(i)}=\{\pi_{0}^{(i)}(1),\dots,\pi_{0}^{(i)}(d_{0})\}\trans$, respectively, 
then the stationary distribution of $\bP_{0,x_{1},\dots,x_{p}}^{(i)}(\cdot \mid y_{t-1})$, 
denoted by $\bpi_{0,x_{1},\dots,x_{p}}^{(i)}=\{\pi_{0,x_{1},\dots,x_{p}}^{(i)}(1),\dots,\pi_{0,x_{1},\dots,x_{p}}^{(i)}(d_{0})\}\trans$, 
has a representation 
$\pi_{0,x_{1},\dots,x_{p}}^{(i)}(y_{t})=\pi_{0}(y_{t})\pi_{0,x_{1},\dots,x_{p}}(y_{t})+\pi_{1}(y_{t})\pi_{0}^{(i)}(y_{t})$. 
Conversely, if $\pi_{0}(y_{t-1}) \in (0,1)$, $\bP_{0,x_{1},\dots,x_{p}}^{(i)}(\cdot \mid y_{t-1})$ can be ergodic even when neither of the two component distributions are so. 
This can be seen by constructing an example with binary state space $\{1,2\}$ 
where one of the component transition distributions only allows self transitions ($1 \to 1, 2 \to 2$) and the other only transitions to the other state ($1 \to 2, 2 \to 1$). 
These results all follow from basic definitions of stationarity and also extend naturally to population level transition distributions $\bP_{0,x_{1},\dots,x_{p}}(\cdot \mid y_{t-1})$.

We now discuss model flexibility, prior support and posterior consistency. 
%
The proposed mixed effect Markov model assumes additivity of predictor effects and individual effects directly on the probability scale. 
The model assumes an implicit upper bound $\pi_{1}(y_{t-1})$ on how far the individual effects $\pi_{1}(y_{t-1})\blambda^{(i)}(y_{t} \mid y_{t-1})$ can stretch 
the effects due to the exogenous predictors $\pi_{0}(y_{t-1})\blambda_{h_{1},\dots,h_{p}}(y_{t} \mid y_{t-1})$ in modeling $\bP_{h_{1},\dots,h_{p}}^{(i)}(y_{t} \mid y_{t-1})$. 
This bound can be easily relaxed by allowing the $\pi_{1}(y_{t-1})$'s to also be individual specific with a multi-level hierarchical prior on them. 
We have not pursued such generalizations in this article in favor of simplicity and parsimony. 
Being based on the partition model for MCEPs introduced in Section \ref{sec: pm for mcep} in the main paper, 
the model for the population level mean transition probabilities 
$\bP_{h_{1},\dots,h_{p}}(\cdot\mid y_{t-1})$, on the other hand, is fully nonparametric, 
taking into account all order interactions between the exogenous and the local predictors.  
The class $\calP_{0}$, defined above, thus denotes a fairly large class of individual specific exogenous predictor dependent transition distributions.  

It is easy to check that our assumed priors, referred to collectively as $\Pi$, 
assign positive probability on any arbitrarily close $L_{1}$ neighborhood of 
any $\bP_{0} = \{P_{0,x_{s,1},\dots,x_{s,p}}^{(i_{s})}(y_{t} \mid y_{t-1})\}_{s=1}^{s_{0}} \in \calP_{0}$. 
More formally, 
with $d(P_{x_{1},\dots,x_{p}}^{(i)},P_{0,x_{1},\dots,x_{p}}^{(i)})=\sum_{y_{t-1}=1}^{d_{0}}\sum_{y_{t}=1}^{d_{0}}\abs{P_{x_{1},\dots,x_{p}}^{(i)}(y_{t} \mid y_{t-1})-P_{0,x_{1},\dots,x_{p}}^{(i)}(y_{t} \mid y_{t-1})}$,
we have 
$\Pi \{\B_{\delta}(\bP_{0})\}>0$ 
for any 
$\bP_{0} \in \calP_{0}$ 
and any $\delta>0$, 
where $\B_{\delta}(\bP_{0})=\{P_{x_{s,1},\dots,x_{s,p}}^{(i_{s})}: d(P_{x_{s,1},\dots,x_{s,p}}^{(i_{s})},P_{0,x_{s,1},\dots,x_{s,p}}^{(i_{s})}) \leq \delta, s=1,\dots,s_{0}\}$.

Let $\calP_{00} \subset \calP_{0}$ be the class of ergodic transition probability distributions $\bP_{0}$
with associated stationary distributions $\pi^{(i)}_{0,x_{1},\dots,x_{p}}(y_{t})$, where $\pi^{(i)}_{0,x_{1},\dots,x_{p}}(y_{t})>0$ for all $y_{t}\in \S_{0}$. 
Assuming $\{y_{s,t}\}_{s=1,t=1}^{s_{0},T_{s}}$ to be ergodic with the true transition dynamics characterized by some $\bP_{0}\in \calP_{00}$, 
it then follows, using strong law of large numbers for ergodic Markov chains (Eichelsbacher and Ganesh, 2002), 
that the posterior $\Pi[ \cdot \mid \{x_{s,j},y_{s,t}\}_{s=1,t=1,j=1}^{s_{0},T_{s},p}]$ 
concentrates almost surely in arbitrarily small neighborhoods of the true data generating parameters $\bP_{0}$ as $\min_{s}T_{s} \to \infty$ 
(Ghosh and Ramamoorthi, 2003).  
Formally, for any $\delta>0$ and any $\bP_{0} \in \calP_{00}$, 
$\Pi[\B_{\delta}(\bP_{0}) \mid \{x_{s,j},y_{s,t}\}_{s=1,t=1,j=1}^{s_{0},T_{s},p}] \to 1~\text{almost surely}~\bP_{0}~\text{as}~\min_{s}  T_{s}\to \infty$.
%

Asymptotic regimes for classical mixed effects models typically assume the number of subjects to approach infinity while the number of observations for each subject remains fixed. 
The criteria considered here, on the contrary, assumes the number of subjects to remain fixed but assumes the length of each sequence to approach infinity. 
This is a more realistic scenario for animal vocalization experiments, 
since it is practically impossible to study more than a small to moderate number of mice from each genotype.
The recording times of the songs, however, can be easily increased. 

\section{Additional Figures}

This section presents additional figures summarizing the analyses of the Foxp2 data set and the simulation experiments, 
described in Sections \ref{sec: afds} and \ref{sec: se} in the main paper, respectively.

\begin{figure}[hp]
\begin{center}
\includegraphics[height=8cm, width=15cm, trim=4cm 1cm 3cm 1cm, clip=true]{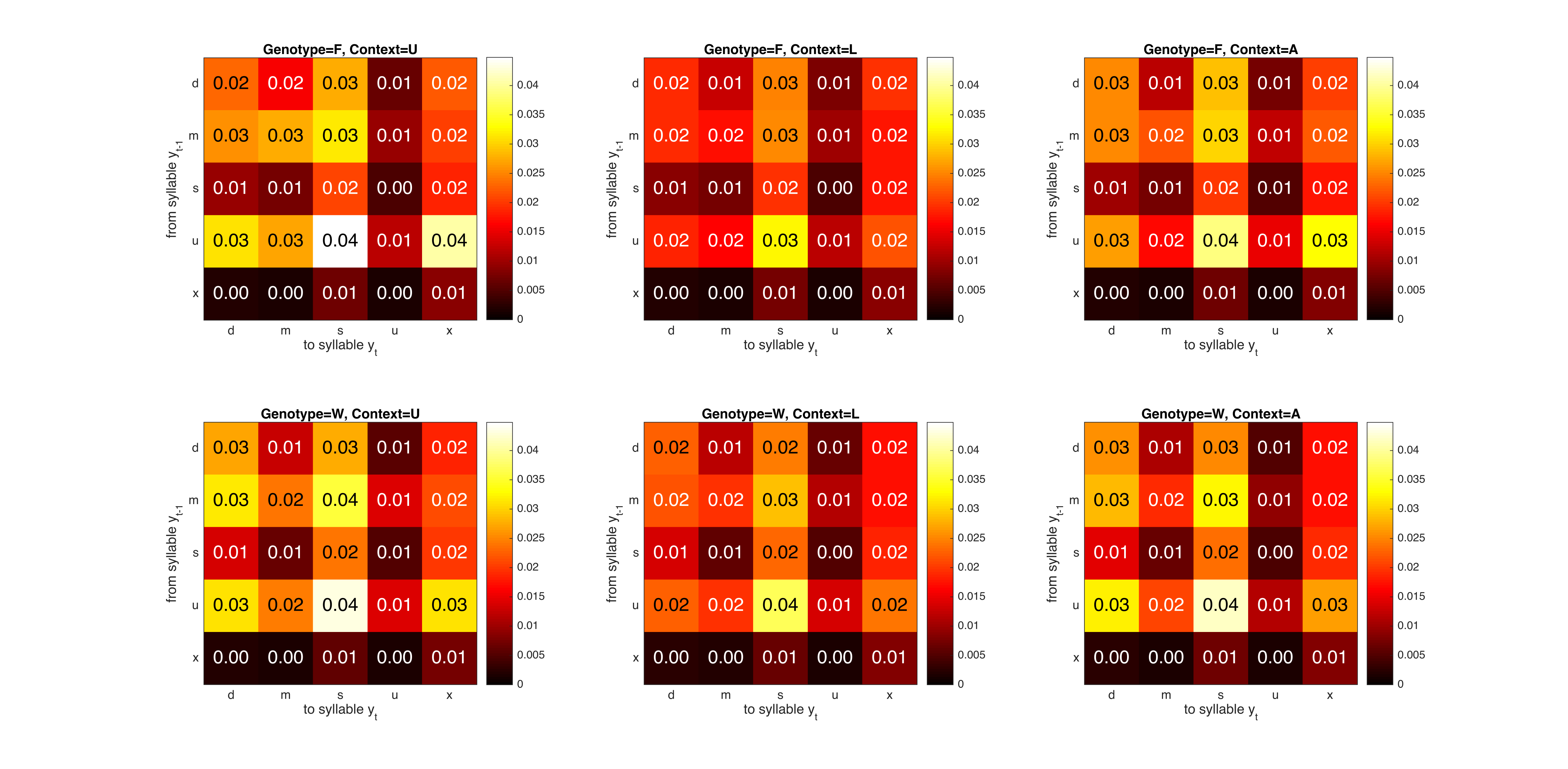}
\caption{Results for the Foxp2 data set. 
Estimated posterior standard deviation of transition probabilities $P_{x_{1},x_{2}}(y_{t} \mid y_{t-1})$ for syllables $y_{t},y_{t-1} \in \{d,m,s,u,x\}$ for different combinations of genotype $x_{1} \in \{F,W\}$ and social contexts $x_{2} \in \{U,L,A\}$.}
\label{fig: Foxp2 post sds}
\vskip 15pt
\includegraphics[height=8cm, width=15cm, trim=4cm 1cm 3cm 1cm, clip=true]{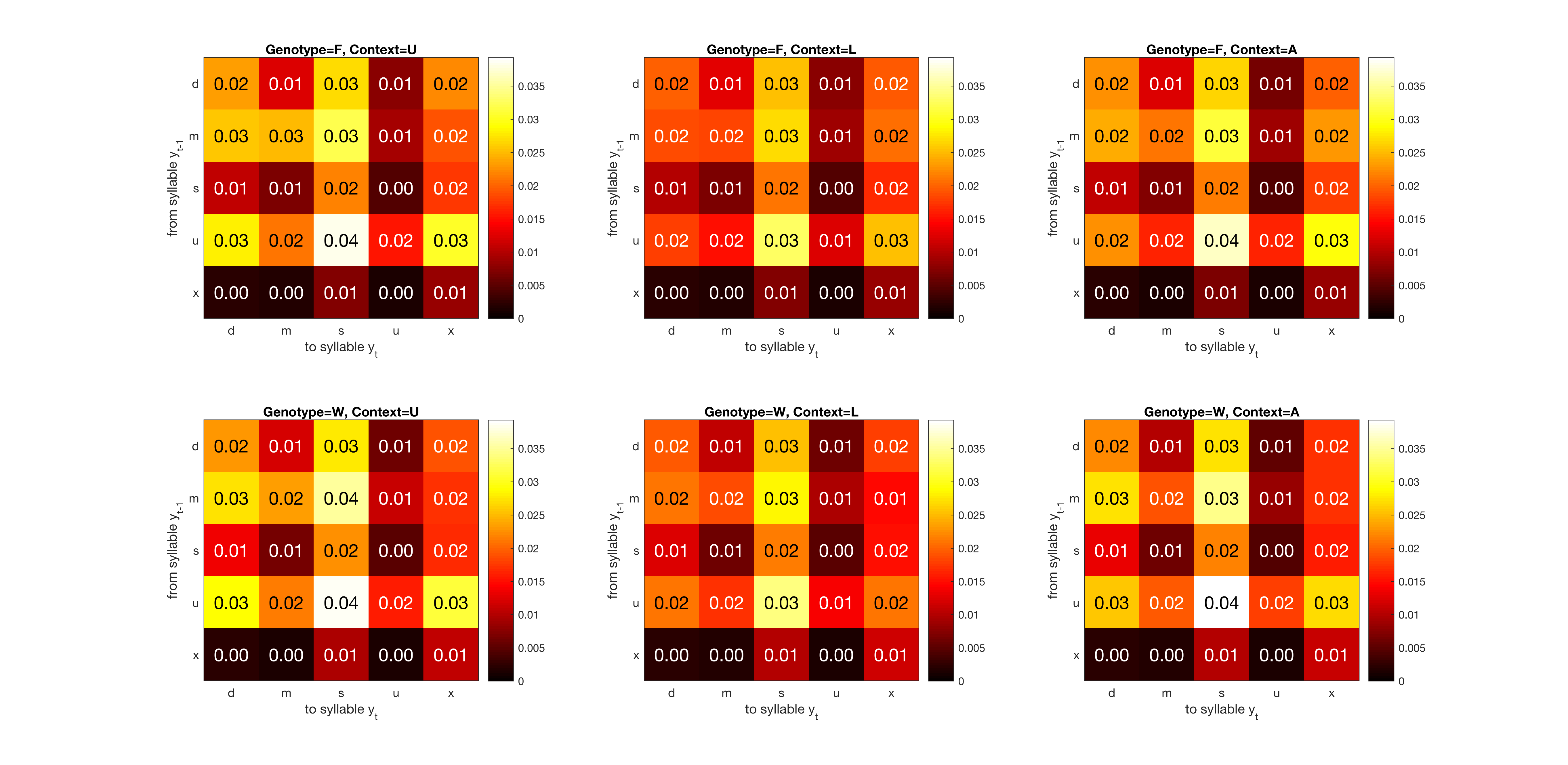}
\caption{Results for the simulation scenario D described in Section \ref{sec: se} of the main paper. 
Estimated posterior standard deviation of transition probabilities $P_{x_{1},x_{2}}(y_{t} \mid y_{t-1})$ for syllables $y_{t},y_{t-1} \in \{d,m,s,u,x\}$ for different combinations of genotype $x_{1} \in \{F,W\}$ and social contexts $x_{2} \in \{U,L,A\}$.}
\label{fig: sim study D post sds}
\end{center}
\end{figure}

\begin{figure}[hp]
\begin{center}
\includegraphics[height=4cm, width=4.5cm, trim=1cm 0cm 1cm 1cm, clip=true]{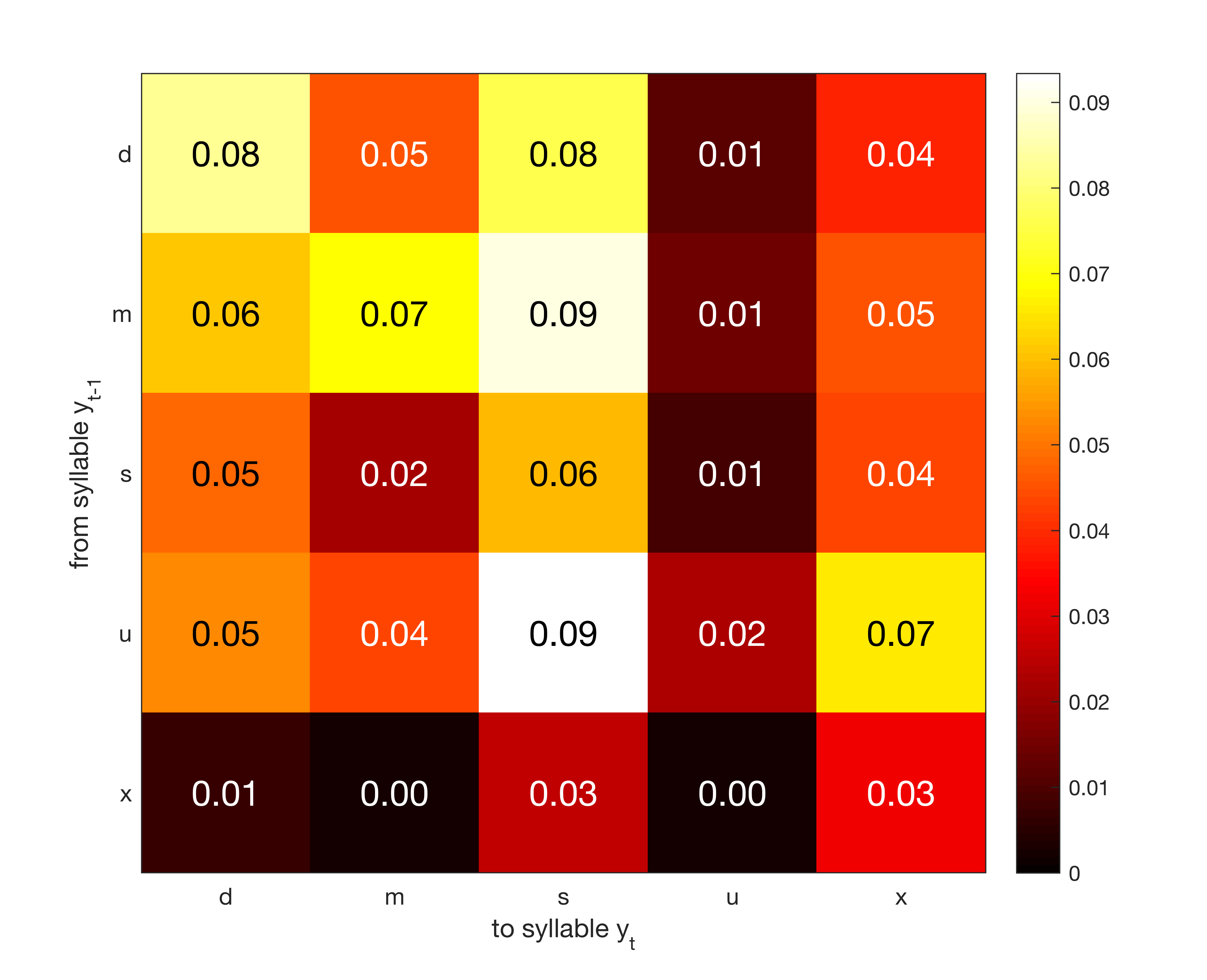}
\caption{Results for the Foxp2 data set. 
Estimated posterior standard deviation of the random effects parameters $\pi_{1}(y_{t-1})\lambda^{(i)}(y_{t} \mid y_{t-1})$ for syllables $y_{t},y_{t-1} \in \{d,m,s,u,x\}$.}
\label{fig: Foxp2 post sds random effects}
\vskip 15pt
\includegraphics[height=4cm, width=4.5cm, trim=1cm 0cm 1cm 1cm, clip=true]{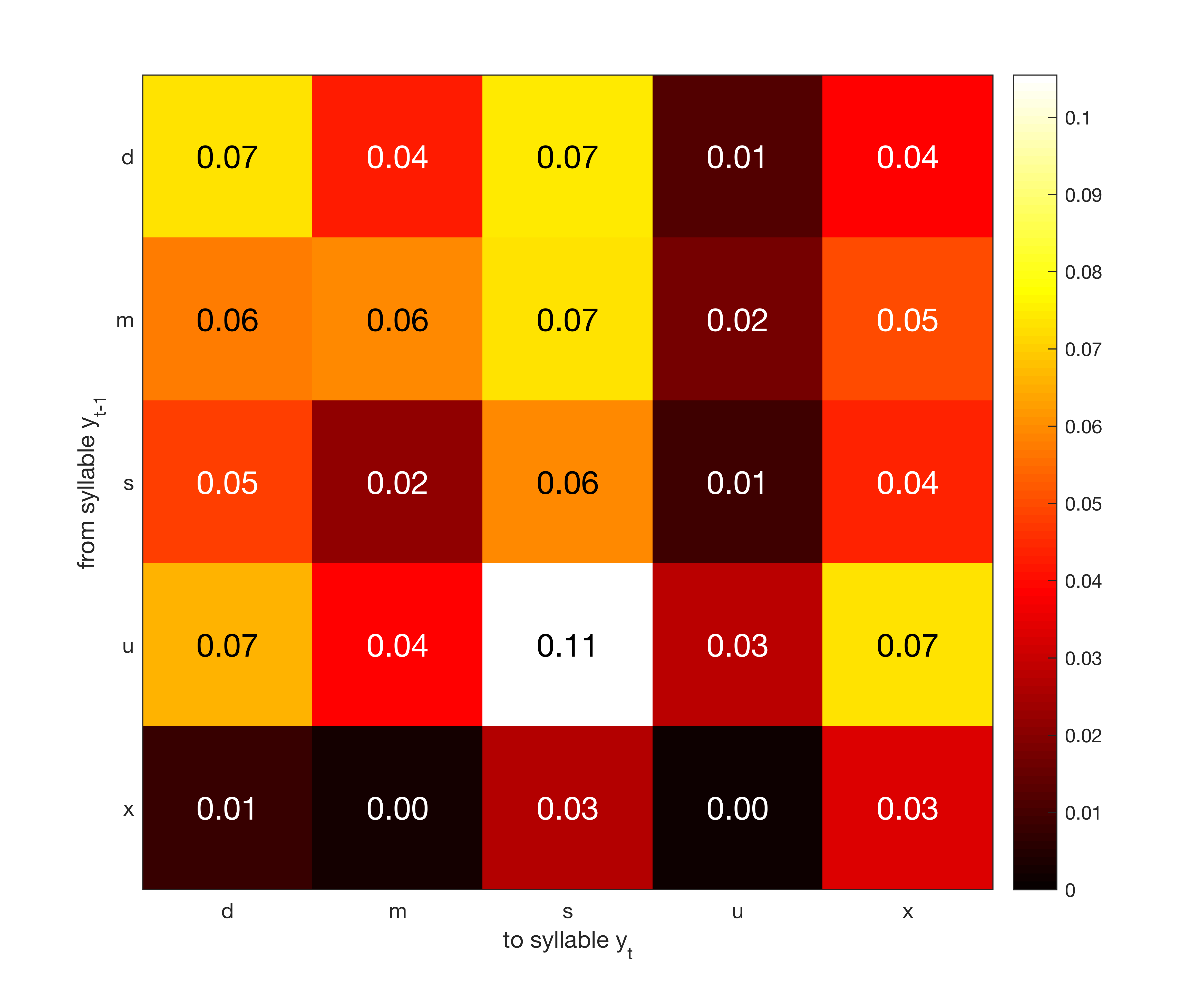}
\caption{Results for the simulation scenario D described in Section \ref{sec: se} of the main paper. 
Estimated posterior standard deviation of the random effects parameters $\pi_{1}(y_{t-1})\lambda^{(i)}(y_{t} \mid y_{t-1})$ for syllables $y_{t},y_{t-1} \in \{d,m,s,u,x\}$.}
\label{fig: sim study D post sds random effects}
\end{center}
\end{figure}

\begin{figure}[hp]
\begin{center}
\includegraphics[height=4.5cm, width=15cm, trim=4cm 0cm 3cm 0cm, clip=true]{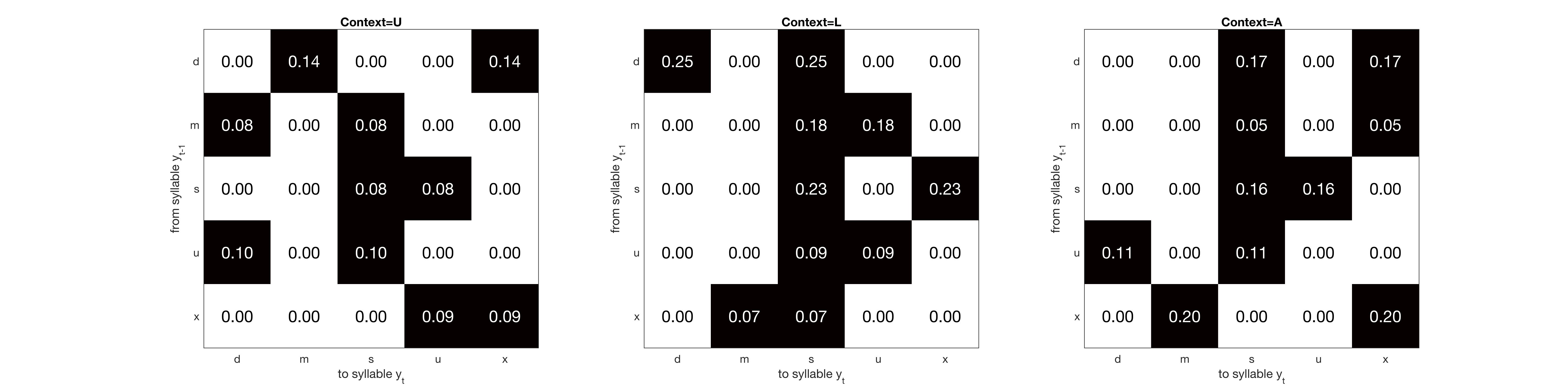}
\includegraphics[height=4.5cm, width=15cm, trim=4cm 0cm 3cm 0cm, clip=true]{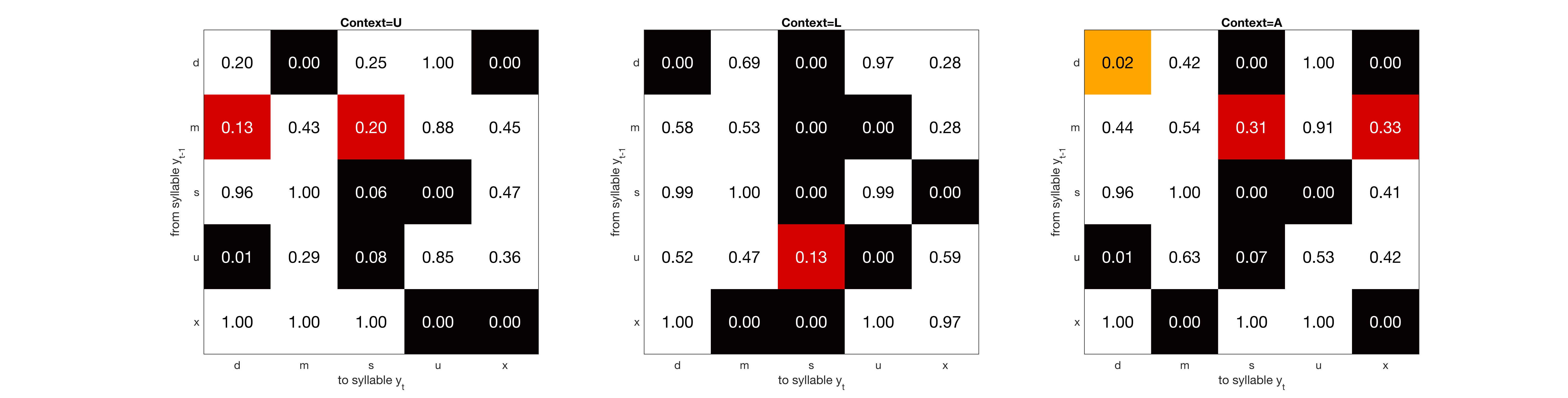}
\includegraphics[height=4.5cm, width=15cm, trim=4cm 0cm 3cm 0cm, clip=true]{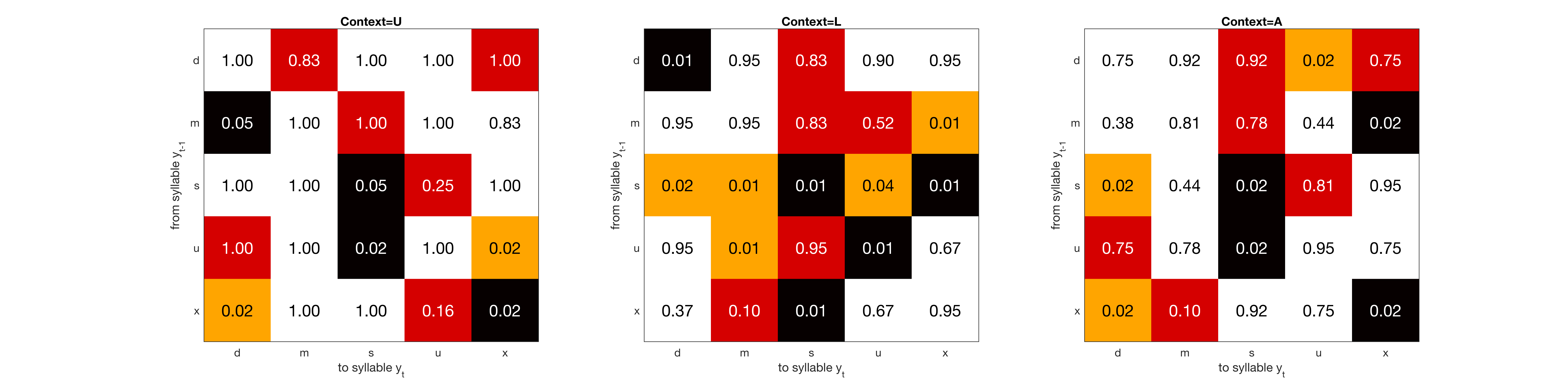}
\caption{Results for the simulation scenario F described in Section \ref{sec: se} in the main paper, but for a different random seed. 
The top row shows the true values of $\abs{\Delta P_{\cdot,x_{2}}(y_{t} \mid y_{t-1})}=\abs{P_{1,x_{2}}(y_{t} \mid y_{t-1}) - P_{2,x_{2}}(y_{t} \mid y_{t-1})}$ 
for syllables $y_{t},y_{t-1} \in \{d,m,s,u,x\}$ and social contexts $x_{2} \in \{U,L,A\}$. 
Positive differences are highlighted in black. 
The middle row shows the estimated posterior probabilities of $H_{0,y_{t} \mid y_{t-1},x_{2}} : \abs{\Delta P_{\cdot,x_{2}}(y_{t} \mid y_{t-1})} \leq 0.02$. 
The bottom row shows Benjamini-Hochberg adjusted p-values obtained using the method of \cite{Chabout_etal:2016}. 
Posterior probabilities smaller than $0.1$ are considered significant and are highlighted in black and orange.
Posterior probabilities greater than $0.1$ are presented in white and red.
Likewise, p-values smaller than $0.1$ are considered significant and are highlighted in black and orange.
P-values greater than $0.1$ are presented in white and red.
White and black cells represent correct decisions, 
orange cells mark rejections of true $H_{0,\ell}$ (false positives), 
and red cells mark failures to reject false $H_{0,\ell}$ (false negatives). 
}
\label{fig: sim study F post prob 2}
\end{center}
\end{figure}

\pagebreak
\section{Comparison with GLM Based Approach} \label{sec: GLM}
In this section, we revisit GLM based approaches to mixed effects Markov chains. 
Adapting to Altman (2007), without any interaction among the local predictor $y_{t-1}$ and the exogenous predictors $x_{j}, j=1,\dots,p$, 
using the logit link, we are now required to formulate $d_{0}-1$ models, one for each $y_{t}=1,\dots,d_{0}-1$, of the form  
\vspace{-2ex}
\bse
&& \log \left\{ \frac{P_{x_{s,1},\dots,x_{s,p}}^{(i_{s})}(y_{s,t} \mid y_{s,t-1})}{P_{x_{s,1},\dots,x_{s,p}}^{(i_{s})}(d_{0} \mid y_{s,t-1})} \right\} = \\
&&~~ \beta_{0,y_{t}}   + \sum_{y_{t-1}=1}^{d_{0}-1}\beta_{y_{t},y_{t-1}} 1\{y_{s,t-1}=y_{t-1}\}  +  \sum_{j=1}^{p}\sum_{x_{j}=1}^{d_{j}-1}\beta_{j,y_{t},x_{j}} 1\{x_{s,j}=x_{j}\} + u_{y_{t},y_{t-1}}^{(i_{s})}, 
\ese
\vspace{-3ex}\\
where $\bu^{(i)}=\{u_{y_{t},y_{t-1}}^{(i)}\}_{y_{t}=1,y_{t-1}=1}^{d_{0}-1,d_{0}}$ are random effects to due to the $i\th$ individual. 
Except for the restrictive special case of binary sequences, estimation of the model parameters becomes prohibitively complex, 
especially in presence of multiple exogenous predictors. 
Incorporating only second order interactions would require an additional 
$N_{int}=\sum_{j_{1}=0}^{p}\sum_{j_{2}=0, j_{1}\neq j_{2}}^{p}(d_{j_{1}}-1)(d_{j_{2}}-1)$ terms for each of the $d_{0}-1$ models, 
significantly increasing model complexities. 
For the Foxp2 application, for instance, this would require $N_{int}=14$ additional terms in each of the $4$ models. 
We have thus ignored interactions among the exogenous and the local predictors here. 

The population average probabilities implied by the model can be obtained by integrating out the random effects as
\vspace{-3ex}
\bse
&&  P_{x_{1},\dots,x_{p}}(y_{t} \mid y_{t-1})  =  \int P_{x_{1},\dots,x_{p}}^{(i)}(y_{t} \mid y_{t-1}) f(\bu_{y_{t-1}}^{(i)})d\bu_{y_{t-1}}^{(i)} = \\
&&~~ \int    \frac{\exp\left(\beta_{0,y_{t}}   +  \beta_{y_{t},y_{t-1}} + \sum_{j=1}^{p} \beta_{j,y_{t},x_{j}} + u_{y_{t},y_{t-1}}^{(i)}\right)}    {\sum_{h=1}^{d_{0}} \exp\left(\beta_{0,h}   +  \beta_{h,y_{t-1}} +  \sum_{j=1}^{p} \beta_{j,h,x_{j}} + u_{h,y_{t-1}}^{(i)}\right)}  f(\bu_{y_{t-1}}^{(i)})d\bu_{y_{t-1}}^{(i)} , 
\ese
\vspace{-3ex}\\
where 
$\beta_{0,d_{0}}=0$, $\beta_{d_{0},y_{t-1}}=\beta_{1,d_{0},y_{t-1}}=\dots=\beta_{p,d_{0},y_{t-1}}=u_{d_{0},y_{t-1}}^{(i)}=0$ for all $y_{t-1}$,
$\bu_{y_{t-1}}^{(i)}=(u_{1,y_{t-1}}^{(i)},\dots,u_{d_{0}-1,y_{t-1}}^{(i)})\trans$, 
and $f(\bu_{y_{t-1}}^{(i)})$ is the random effects distribution. 
Typically it is assumed that $f(\bu_{y_{t-1}}^{(i)}) = \MVN_{d_{0}-1}(\bzero,\bSigma_{u})$, 
where $\MVN_{q}(\bmu,\bSigma)$ denotes a $q$-dimensional multivariate normal distribution with mean vector $\bmu$ and covariance matrix $\bSigma$. 
Often such models are further simplified by assuming $\bu_{y_{t-1}}^{(i)}=\bu^{(i)}$ for all $y_{t-1}$ (Altman, 2007) 
and the components of $\bu_{y_{t}}$ to be distributed independently with $\bSigma_{u}=\diag(\sigma_{u,1}^{2},\dots,\sigma_{u,d_{0}-1}^{2})$.

Even with such restrictive simplifying assumptions, the population level transition probabilities $P_{x_{1},\dots,x_{p}}(y_{t} \mid y_{t-1})$ do not have closed form expressions. 
Assuming $P_{x_{1},\dots,x_{p}}(y_{t} \mid y_{t-1})$ to arise from the same multinomial logit functional form 
\vspace{-3ex}
\bse
P_{x_{1},\dots,x_{p}}(y_{t} \mid y_{t-1}) =  \frac{\exp \left(\beta_{0,y_{t}}^{\star}   +  \beta_{y_{t},y_{t-1}}^{\star} + \sum_{j=1}^{p} \beta_{j,y_{t},x_{j}}^{\star}\right)}    {\sum_{h=1}^{d_{0}} \exp\left(\beta_{0,h}^{\star}   +  \beta_{h,y_{t-1}}^{\star} +  \sum_{j=1}^{p} \beta_{j,h,x_{j}}^{\star} \right)},
\ese
\vspace{-3ex}\\
an approximation yields $\beta_{0,y_{t}}^{\star} \approx \beta_{0,y_{t}}/(1+c^{2}\sigma_{u,y_{t}}^{2})^{1/2}, \beta_{y_{t},y_{t-1}}^{\star} \approx \beta_{y_{t},y_{t-1}}/(1+c^{2}\sigma_{u,y_{t}}^{2})^{1/2}$ and so on, where $c=(16\sqrt{3})/(15\pi)$ (Zeger \emph{et al}., 1988).
Individual and population level fixed effects parameters are thus different and have to be differently interpreted. 
Specifically, population level probabilities depend on individual heterogeneity - 
two populations with different individual heterogeneity will have different population level probabilities even if they have the same individual level fixed effects parameters. 

Testing scientific hypotheses related to influences of the predictors using such GLM based models is also complicated. 
For instance, the global null $H_{0j}$ of no effect of the $j\th$ exogenous predictor $x_{j}$, 
when translated in terms of the model parameters, 
becomes a complicated composite hypothesis $H_{0j}: \beta_{j,y_{t},x_{j}}=0$ for all $y_{t}=1,\dots,d_{0}-1$ and all $x_{j}=1,\dots,d_{j}-1$.

In comparison, our model is highly flexible, 
parsimoniously accommodating interactions of all orders between the exogenous and the local predictors. 
The random effects in our model for the individual level transition probabilities 
can be easily integrated out to obtain closed form expressions of the population level transition probabilities. 
The fixed effects components remain the same in both individual and population level probabilities and hence can be interpreted in the same way. 
Finally, testing scientific hypotheses related to global influences of the predictors is very straightforward using our approach 
as they can be translated in terms of a single model parameter.

\begin{figure}[hp]
\begin{center}
\includegraphics[height=8cm, width=15cm, trim=4cm 1cm 3cm 1cm, clip=true]{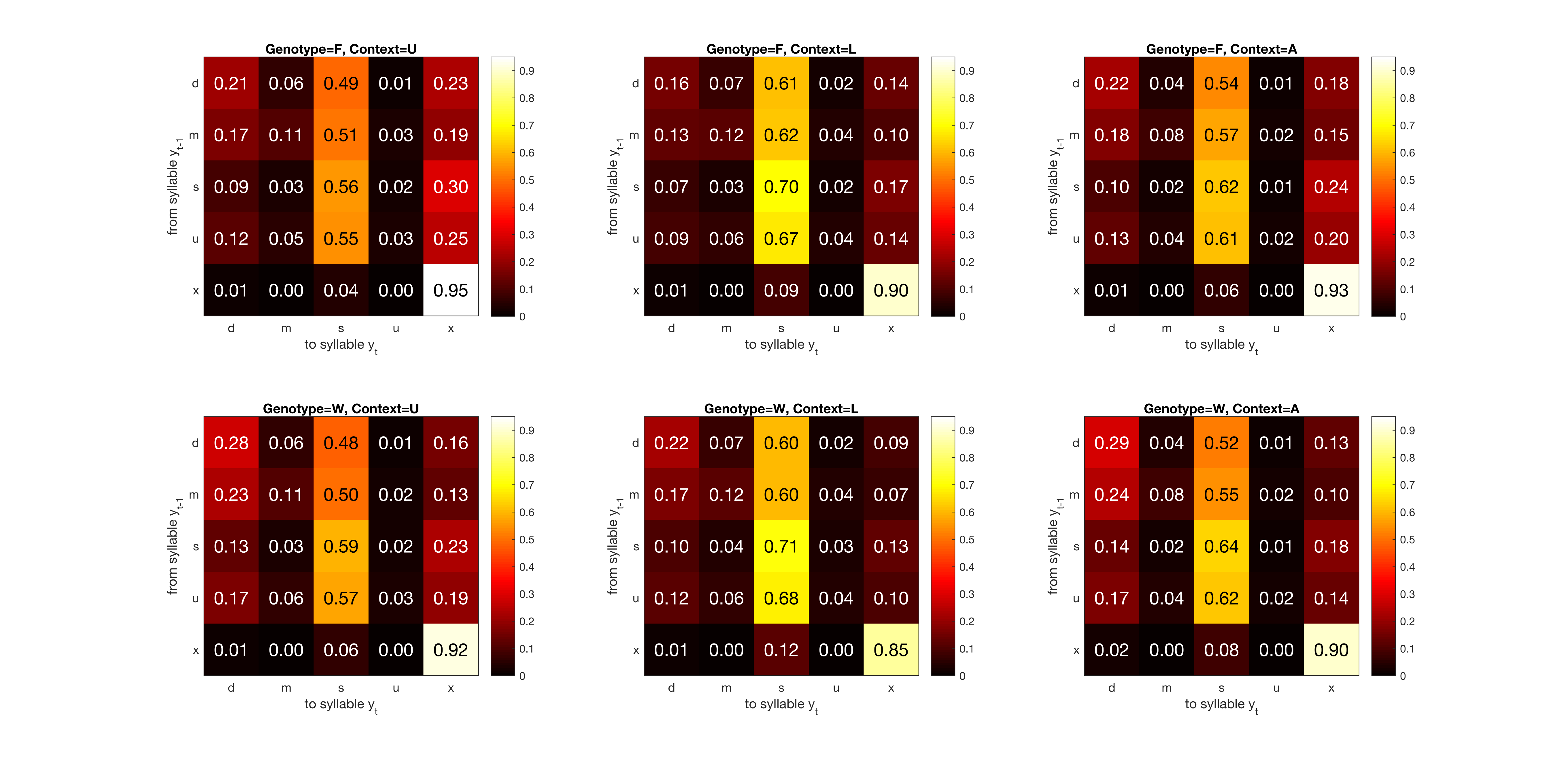}
\caption{Results for the Foxp2 data set for the GLM based approach described in Section \ref{sec: GLM} in the Supplementary Materials. 
Estimated approximate posterior mean transition probabilities $P_{x_{1},x_{2}}(y_{t} \mid y_{t-1})$ for syllables $y_{t},y_{t-1} \in \{d,m,s,u,x\}$ for different combinations of genotype $x_{1} \in \{F,W\}$ and social contexts $x_{2} \in \{U,L,A\}$.}
\label{fig: Foxp2 post mean MCMCglmm}
\vskip 15pt
\includegraphics[height=4cm, width=15cm, trim=4cm 0cm 2cm 0cm, clip=true]{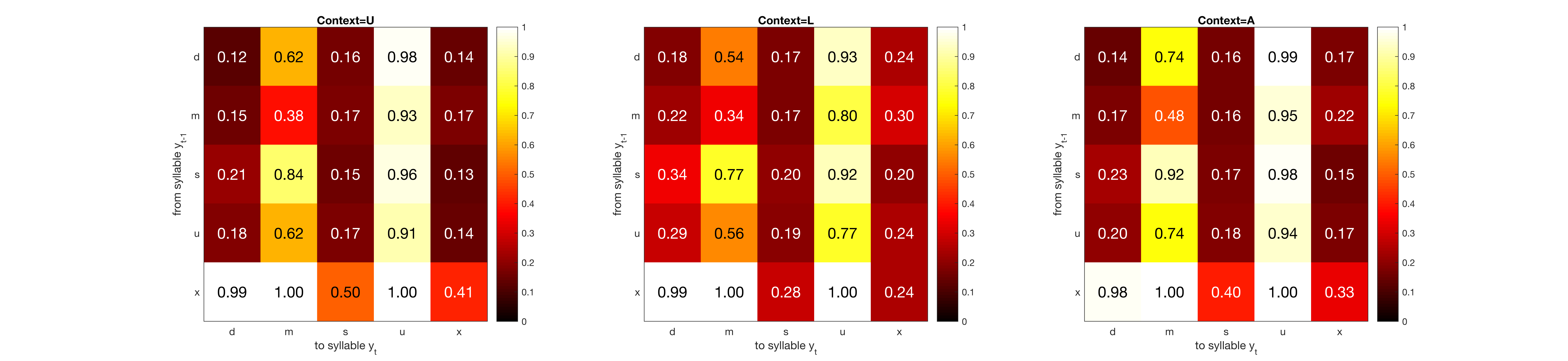}
\caption{Results for the Foxp2 data set for the GLM based approach described in Section \ref{sec: GLM} in the Supplementary Materials. 
The estimated posterior probability of $H_{0,y_{t} \mid y_{t-1},x_{2}} : \abs{\Delta P_{\cdot,x_{2}}(y_{t} \mid y_{t-1})} = \abs{P_{1,x_{2}}(y_{t} \mid y_{t-1}) - P_{2,x_{2}}(y_{t} \mid y_{t-1})} \leq 0.02$.}
\label{fig: Foxp2 post prob MCMCglmm}
\end{center}
\end{figure}

We implemented the multinomial logit based mixed effects Markov model described above using the MCMCglmm package in R (Hadfield, 2010). 
Maximum likelihood estimation of the model parameters 
using other R packages did not produce realistic results. 
Figure \ref{fig: Foxp2 post mean MCMCglmm} shows the estimated posterior means of the population level transition probabilities 
based on $12,000$ samples drawn from the posterior, thinned by an interval of $10$ after the initial $2,000$ were discarded as burnin. 
Comparison with estimates produced by our method, summarized in Figure \ref{fig: Foxp2 post mean} in the main paper, suggests overall agreement. 
For reasons detailed above, global significance of the exogenous predictors could not be straightforwardly assessed. 
We could, however, assess the significance of each $\beta$ parameter from the MCMC output using 
the minimum of the proportion of samples in which $\beta$ is on one side or the other of zero, 
referred to as pMCMC in MCMCglmm. 
For the Foxp2 data set, the four $\beta$ parameters associated with genotype, 
namely $\beta_{1,1,1}, \beta_{1,2,1}, \beta_{1,3,1},\beta_{1,4,1}$, had pMCMC values $0.446$, $0.436$, $0.368$ and $0.106$, 
indicating none of them to be marginally significant.   
To assess local differences in transition probabilities between the two genotypes, 
we employed the approach developed in Section \ref{sec: testing} of the main paper. 
Figure \ref{fig: Foxp2 post prob MCMCglmm} summarizes 
the posterior probabilities of the local null hypotheses 
$H_{0,y_{t} \mid y_{t-1},x_{2}} : \abs{\Delta P_{\cdot,x_{2}}(y_{t} \mid y_{t-1})} \leq 0.02$ estimated from the MCMC output of the GLM based model. 
Unlike the results produced by our approach, summarized in Figure \ref{fig: Foxp2 post prob} in the main paper, 
no local difference was found to be significant at the $0.10$ posterior probability level.

\begin{figure}[hp]
\begin{center}
\includegraphics[height=4.5cm, width=15cm, trim=4cm 0cm 3cm 0cm, clip=true]{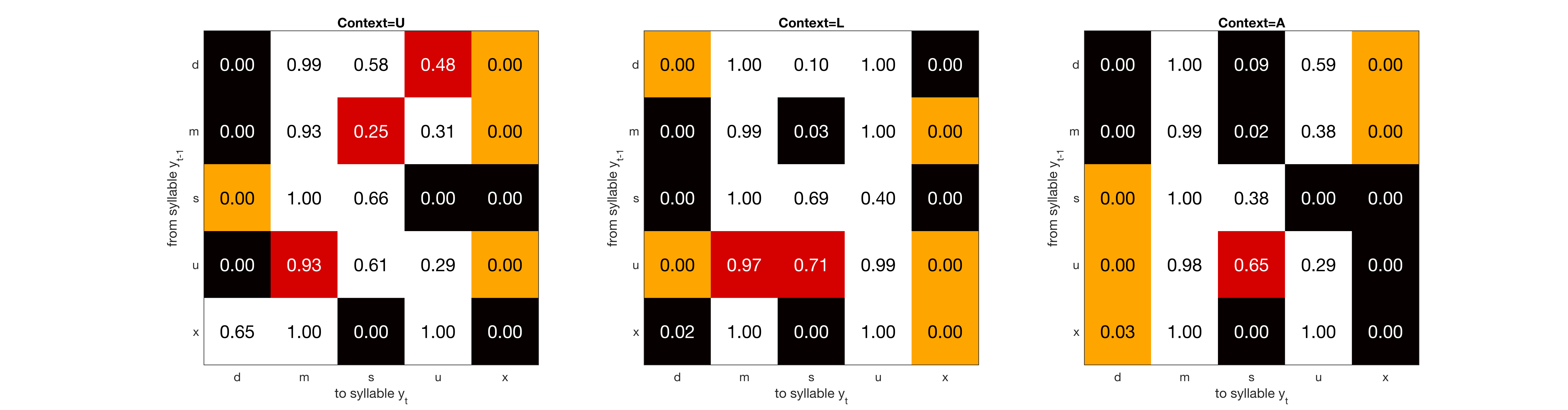}
\caption{Results for the simulation scenario F described in Section \ref{sec: se} in the main paper. 
These results were produced by the GLM based approach described in Section \ref{sec: GLM} in the Supplementary Materials. 
The results show the estimated posterior probabilities of $H_{0,y_{t} \mid y_{t-1},x_{2}} : \abs{\Delta P_{\cdot,x_{2}}(y_{t} \mid y_{t-1})} \leq 0.02$. 
Posterior probabilities smaller than $0.1$ are considered significant and are highlighted in black and orange.
Posterior probabilities greater than $0.1$ are presented in white and red.
White and black cells represent correct decisions, 
orange cells mark rejections of true $H_{0,\ell}$ (false positives), 
and red cells mark failures to reject false $H_{0,\ell}$ (false negatives). 
Compare with Figure \ref{fig: sim study F post prob} in the main paper. 
}
\label{fig: sim study F post prob MCMCglmm}
\end{center}
\end{figure}

\begin{figure}[hp]
\begin{center}
\includegraphics[height=4.5cm, width=15cm, trim=4cm 0cm 3cm 0cm, clip=true]{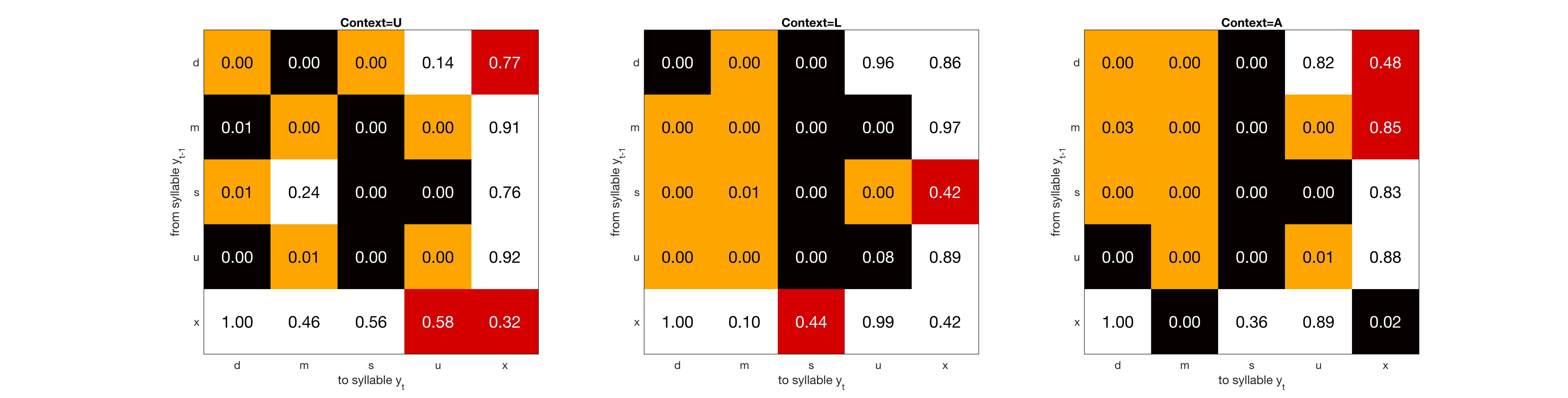}
\caption{Results for the simulation scenario F described in Section \ref{sec: se} in the main paper, but for a different random seed. 
These results were produced by the GLM based approach described in Section \ref{sec: GLM} in the Supplementary Materials. 
The results show the estimated posterior probabilities of $H_{0,y_{t} \mid y_{t-1},x_{2}} : \abs{\Delta P_{\cdot,x_{2}}(y_{t} \mid y_{t-1})} \leq 0.02$. 
Posterior probabilities smaller than $0.1$ are considered significant and are highlighted in black and orange.
Posterior probabilities greater than $0.1$ are presented in white and red.
White and black cells represent correct decisions, 
orange cells mark rejections of true $H_{0,\ell}$ (false positives), 
and red cells mark failures to reject false $H_{0,\ell}$ (false negatives). 
Compare with Figure \ref{fig: sim study F post prob 2} in the Supplementary Materials. 
}
\label{fig: sim study F post prob 2 MCMCglmm}
\end{center}
\end{figure}

To further assess how the multinomial logit based mixed effects Markov model compares with our proposed approach in detecting local differences in transition probabilities between the two genotypes, 
we compared the results produced by the two methods for data sets simulated under scenario F described in Section \ref{sec: se} in the main paper. 
The posterior means of the population level transition probabilities estimated by the GLM based approach (not shown here)
were quite different from the truth.  
Figures \ref{fig: sim study F post prob MCMCglmm} and \ref{fig: sim study F post prob 2 MCMCglmm} summarize the estimated posterior probabilities of the local null hypotheses 
$H_{0,y_{t} \mid y_{t-1},x_{2}} : \abs{\Delta P_{\cdot,x_{2}}(y_{t} \mid y_{t-1})} \leq 0.02$ for two different simulated data sets. 
Compared to the results produced by our approach, summarized in Figure \ref{fig: sim study F post prob} and Figure \ref{fig: sim study F post prob 2}, 
there were many more false decisions. 

\vspace{0.5cm}

\newpage
\section*{References}
\refmark
Altman, R. M. (2007). Mixed hidden Markov models: an extension of the hidden Markov model to the longitudinal data setting. 
\emph{Journal of the American Statistical Association}, {\bf{102}}, 201-210.

\refmark
Eichelsbacher, P. and Ganesh, A. (2002). Bayesian inference for Markov chains. \emph{Journal of Applied Probability}, {\bf{39}}, 91-99.

\refmark
Ghosh, J. K. and Ramamoorthi, R. V. (2003). \emph{Bayesian nonparametrics}. Springer Verlag, Berlin.

\refmark
Hadfield, J. D. (2010). MCMC methods for multi-response generalized linear mixed models: The MCMCglmm R package. 
\emph{Journal of Statistical Software}, {\bf{33}}, 1-22.

\refmark
Zeger, S. L., Liang, K.-Y., and Albert, P. S. (1988). Models for longitudinal data: a generalized estimating equation approach. 
\emph{Biometrics}, {\bf{44}}, 1049-1060.

\end{document}